\documentclass[useAMS,usenatbib]{mn2e}

\usepackage{graphicx}
\usepackage{amssymb}
\usepackage{amsmath}

\title[Locating positions of $\gamma$-ray--emitting regions]
{Locating positions of $\gamma$-ray--emitting regions in blazars}
\author[H. T. Liu, J. M. Bai and J. M. Wang]{H. T. Liu$^{1,2}$\thanks{E-mail:
htliu@ynao.ac.cn; baijinming@ynao.ac.cn; wangjm@mail.ihep.ac.cn},
J. M. Bai$^{1,2}$\footnotemark[1] and
J. M. Wang$^{3,4}$\footnotemark[1]\\
$^{1}$National Astronomical Observatories/Yunnan Astronomical
Observatory, Chinese Academy of Sciences, \\ Kunming,
Yunnan 650011, China\\
$^{2}$Key Laboratory for the Structure and Evolution of Celestial
Objects, Chinese Academy of Sciences, \\ Kunming, Yunnan 650011,
China\\
$^{3}$Key Laboratory for Particle Astrophysics, Institute of High
Energy Physics, Chinese Academy of Sciences,\\ 19B Yuquan
Road, Beijing 100049, China\\
$^{4}$Theoretical Physics Center for Science Facilities, Chinese
Academy of Sciences, Beijing 100049, China}

\begin{document}

\date{Accepted . Received }


\maketitle

\label{firstpage}

\begin{abstract}
We propose a new method to locate the $\gamma$-ray--emitting
positions $R_{\rm{\gamma}}$ from the measured time lags
$\tau_{\rm{ob}}$ of $\gamma$-ray emission relative to broad
emission lines. The method is also applicable to lower
frequencies. $R_{\rm{\gamma}}$ depends on parameters
$\tau_{\rm{ob}}$, $R_{\rm{BLR}}$, $v_{\rm{d}}$ and $\theta$, where
$R_{\rm{BLR}}$ is the size of broad-line region, $v_{\rm{d}}$ is
the travelling speed of disturbances down the jet and $\theta$ is
the viewing angle of the jet axis to the line of sight. As
$\tau_{\rm{ob}}=0$, $\tau_{\rm{ob}}<0$ or $\tau_{\rm{ob}}>0$, the
broad lines zero-lag, lag or lead the $\gamma$-rays, respectively.
It is applied to 3C 273, in which the lines and the radio emission
have enough data, but the $\gamma$-rays have not. We find
$\tau_{\rm{ob}}<0$ and $\tau_{\rm{ob}}>0$ for the 5, 8, 15, 22 and
37 GHz emission relative to the broad lines H$\alpha$, H$\beta$
and H$\gamma$. The lag may be positive or negative, however
current data do not allow to discriminate between the two cases.
The measured lags are on the order of years. For a given line,
$\tau_{\rm{ob}}$ generally decreases as radio frequency increases.
This trend most likely results from the radiative cooling of
relativistic electrons. The negative lags have an average of
$\tau_{\rm{ob}}=-2.86$ years for the 37 GHz emission, which
represents that the lines lag the radio emission. The positive
lags have $\tau_{\rm{ob}}=3.20$ years, which represents that the
lines lead the radio emission. We obtain the radio emitting
positions $R_{\rm{radio}}=0.40$--2.62 pc and
$R_{\rm{radio}}=9.43$--62.31 pc for the negative and positive
lags, respectively. From the constraint of $R_{\rm{\gamma}}\la
R_{\rm{radio}}$ \citep[e.g.][]{b24,b41}, we have
$R_{\rm{\gamma}}\la 0.40$--2.62 pc for the negative lags. For the
positive lags, 4.67--$30.81<R_{\rm{\gamma}}\la 9.43$--62.31 pc.
These estimated $R_{\rm{\gamma}}$ are consistent with those of
other researches. These agreements confirm the reliability of the
method and assumptions. The method may be also applicable to BL
Lacertae objects, in which broad lines were detected.
\end{abstract}


\section{Introduction}
The $\gamma$-rays of blazars are generally believed to be
generated by inverse Compton (IC) emission from a relativistic jet
oriented at a small angle to the line of sight \citep{b14}. Two of
the most accepted scenarios for broad-band emission from radio to
$\gamma$-rays are the synchrotron self-Compton (SSC) and external
Compton (EC) models \citep[see e.g.][]{b34}. The broad-band
spectral energy distributions (SEDs) of blazars consist of two
broad bumps \citep[see e.g.][]{b29,b34}. The first component is
from the synchrotron process, and the second one, generally
peaking at the $\gamma$-ray regime, is generated by the IC
emission of the same electron population responsible for the
synchrotron emission \citep[see e.g.][]{b34,b17}. However, the
positions of $\gamma$-ray--emitting regions are still an open and
controversial issue in the researches on blazars. It was suggested
that $\gamma$-rays are produced within broad-line region (BLR) and
that the $\gamma$-ray--emitting positions $R_{\gamma}$ range
roughly between 0.03 and 0.3 parsec (pc) \citep{b33}. \citet{b11}
also suggested a sub-pc $\gamma$-ray--emitting region. It was
argued that the radiative plasma in relativistic jets of powerful
blazars are inside the BLR \citep{b32}. On the contrary, it was
also argued that the $\gamma$-ray--emitting regions are outside
the BLR \citep{b54,b79}. Internal absorption for 10 GeV--1 TeV
$\gamma$-rays were used to constrain $R_{\gamma}$
\citep{b55,b56,b10}. Variability of the high energy flux indicates
that the $\gamma$-ray--emitting positions cannot be too distant
from the central supermassive black hole \citep{b33,b36}, while
the photon-photon absorption implies that the emitting positions
cannot be too close to the black hole and its accretion disc
\citep{b33,b93,b55,b56,b78,b10,b36,b82}. Bracketed by the two
limits, one obtains a few hundreds of Schwarzschild radii as the
preferred jet location where most of the dissipation occurs
\citep{b36}. \citet{b41} established a connection between
ejections of superluminal radio knots and $\gamma$-ray outbursts
observed by EGRET. They concluded that the radio and $\gamma$-ray
events are originating from the same region of a relativistic jet.
\citet{b91} found that the blazar $\gamma$-ray emission might
depend on the mass of the central black hole and that very high
energy (VHE) $\gamma$-ray--emitting active galactic nuclei (AGNs)
have the black hole masses larger than $10^8$ solar masses.
\citet{b16} confirmed the radio and $\gamma$-ray correlation of
EGRET blazars. \citet{b83} reported that a central feature of the
EGRET results is the high degree of variability seen in many
$\gamma$-ray sources, indicative of the powerful central engines
at work in objects visible to $\gamma$-ray telescopes.

The operation of {\it Fermi Gamma Ray Space Telescope} ({\it
Fermi\/})--Large Area Telescope (LAT) \citep[see e.g.][]{b1,b2}
presents an exceptional opportunity for understanding the central
engines operating in blazars. The First LAT AGN Catalog includes
709 AGNs, comprising 300 BL Lacertae objects (BL Lacs), 296 flat
spectrum radio quasars (FSRQs), 41 AGNs of other types and 72 AGNs
of unknown types \citep{b2}. The distribution of $\gamma$-ray
photon spectral index is found to correlate strongly with blazar
subclass \citep{b2,b4}. \citet{b3} found for {\it Fermi\/} blazars
that variation amplitudes are larger for FSRQs and
low/intermediate synchrotron frequency peaked BL Lacs. \citet{b38}
obtained a few hundreds of Schwarzschild radii as the preferred
jet dissipation region. \citet{b37} found for bright {\it Fermi\/}
blazars that the jet dissipation region is within the BLR for
FSRQs. \citet{b50} identified the pc-scale radio core as a likely
location for both the $\gamma$-ray and radio flares. \citet{b77}
suggested that the blazar emission zone is located at pc-scale
distances from the nucleus, and they also proposed that the
pc-scale blazar activity can be occasionally accompanied by
dissipative events taking place at sub-pc distances. \citet{b35}
argued that the bulk of the most luminous blazars already detected
by {\it Fermi\/} should be characterized by large black hole
masses, around $10^9$ solar masses. \citet{b50} found for {\it
Fermi} blazars that the $\gamma$-ray photon flux correlates with
the compact radio density flux.

Disturbances in the central engine are likely transported down the
relativistic jets. This was supported by observations that dips in
the X-ray emission are followed by ejections of bright
superluminal knots in the radio jets of AGNs and microquasars
\citep[e.g.][]{b58,b22,b7}. The events in the central engine,
where the X-rays are produced, will have a direct effect on the
events in the radio jets \citep[e.g.][]{b58,b22}. The disc-jet
connection was suggested by correlations of emission line
luminosity and radio power of jets for various samples of AGNs
\citep{b69,b28,b21,b73,b19,b20,b94,b40}. The relativistic jets can
be ejected from inner accretion disc in the vicinity of the
central black hole \citep[see e.g.][]{b63,b15,b13,b60}. The close
connection between accretion disc and jets indicates that the
central disturbances are likely transported down the jets. For
example, the central disturbances that drive variations of the
central ionizing continuum may be transported with the outward
Alfv\'en waves \citep[see e.g.][]{b60,b47}. By magnetohydrodynamic
mechanisms \citep{b52}, the transported disturbances may be
converted into the local energies of plasma far from the central
black hole. Thus the disturbances in the central engine would
influence the $\gamma$-rays emitted by the relativistic jet
aligned with the line of sight. At the same time, these
disturbances can lead to the variations of the central ionizing
continuum that drives broad emission lines from BLR. Hence, the
broad lines from the BLR and the $\gamma$-rays from the
relativistic jet may be both coupled to the disturbances in the
central engine.

Based on the photoionization assumption and the time lags between
broad lines and continuum, reverberation mapping observations are
able to determine the sizes of the BLRs for type 1 AGNs \citep[see
e.g. ][]{b45,b92,b46,b44,b43,b64,b65,b66,b89}. According to the
reverberation mapping model \citep[e.g.][]{b12}, the variations of
broad lines can reflect the disturbances in the central engines of
blazars, even though the beamed emission from the relativistic jet
strongly affects the real ionizing continuum from accretion disc
so that no appropriate continuum could be used as the reference to
estimate the time lags relative to the broad lines. If the broad
lines from the BLR and the $\gamma$-rays from the relativistic jet
are both coupled to the disturbances in the central engine, the
disturbances should similarly influence both variations of the
$\gamma$-rays and the broad lines. Thus there should be
correlations and time lags between the broad lines and the
$\gamma$-rays. The time lags should be related to $R_{\gamma}$. In
the paper, we attempt to locate $R_{\gamma}$ from the time lags
between both variations of the broad lines from the BLR and the
$\gamma$-rays from the relativistic jet.

\section{Method}
The broad lines from the BLR and the $\gamma$-rays from the
relativistic jet may be both coupled to the disturbances in the
central engine. Thus the disturbances could similarly influence
both variations of the broad lines and the $\gamma$-rays emitted
by the jet aligned with the line of sight. The outbursts seen in
light curves are physically linked to the ejections of
superluminal radio knots \citep[e.g.][]{b86}. The events in the
central engine will have a direct effect on the events in the
radio jets \citep[e.g.][]{b58,b22}. Thus it is likely that the
$\gamma$-ray outbursts are caused by the disturbances from the
central engine, but not the local disturbances produced in the
jets. Hence, there should be correlations between the $\gamma$-ray
outbursts and the variations of broad lines. It is possible that
there are time lags in the correlations and the time lags are
related to the positions $R_{\rm{\gamma}}$. Thus $R_{\rm{\gamma}}$
could be located by the time lags between the $\gamma$-ray
outbursts and the variations of broad lines. In the future, the
quasi-simultaneous observations of $\gamma$-rays with {\it
Fermi}/LAT and broad lines with optical telescopes on the order of
years may be employed to test this expectation. The method of
locating $R_{\rm{\gamma}}$ is also applicable to infrared, optical
and radio emission.

First, it is assumed that the disturbances in the central engine
are transported outward by some process, and the disturbances
could similarly influence both variations of the broad lines from
the BLR and the $\gamma$-rays from the relativistic jet aligned
with the line of sight. Second, we assume a simple geometry (see
Fig. 1$a$) that is similar to the schemes in the classical
reverberation mapping of the broad lines \citep[see
e.g.][]{b45,b92,b46,b44,b43,b64,b65,b66,b89}. The differences
between this method and the classical mapping are as follows. In
the classical mapping, a part of ionizing continuum drives the
broad lines from the BLR, and another directly reaches observers.
Thus the broad lines lag the continuum due to light travelling
time effects, and the time lag corresponds to the size of the BLR.
In the method proposed here, the ionizing photon signals directly
detected by telescopes in the classical mapping are replaced with
the transporting disturbances from the central engine down the jet
and then with the jet emission signals at $R_{\rm{\gamma}}$. Thus
the method would be valid even if the BLR has a complex
configuration, such as a spherical shell. For simplicity for
graphic illustration, we choose the BLR to be a ring (see Fig.
1$a$). The transporting speed of disturbances down the jet cannot
exceed the speed of light.

Because $R_{\rm{\gamma}}$ is unknown, the $\gamma$-rays may lag,
lead or zero-lag the broad lines. First, we try to find the
position where the $\gamma$-rays are produced and zero-lag the
broad lines. As the disturbances reach point $G$ (see Fig. 1$a$),
where the $\gamma$-rays are produced, i.e. $R_{\rm{\gamma}}=AG$,
the ionizing continuum photons travel from point $A$ to $B$ and
the line photons travel from point $B$ to $I$ in time interval
$R_{\rm{\gamma}}/v_{\rm{d}}$. In the case of zero-lag (hereafter
Case A), we have
$(R_{\rm{BLR}}+R_{\rm{\gamma}})/c=R_{\rm{\gamma}}/v_{\rm{d}}$, and
then
\begin{equation}
R_{\rm{\gamma}}=R_{\rm{BLR}} \frac{v_{\rm{d}}}{c-v_{\rm{d}}},
\end{equation}
where $v_{\rm{d}}$ is the travelling speed of disturbances down
the jet and $c$ is the speed of light. For Case A,
$R_{\rm{\gamma}}> R_{\rm{BLR}}$.

Within segment $AG$, the lines will lag the $\gamma$-rays. As the
disturbances reach point $F$, where the $\gamma$-rays are
produced, i.e. $R_{\rm{\gamma}}=AF$, the ionizing continuum
photons reach point $E$. The light travelling time effects for the
ionizing photons from point $E$ to $B$ and the line photons from
$B$ to $K$ result in the time lag of the lines relative to the
$\gamma$-rays (see Fig. 1$a$). In this case (hereafter Case B), we
have
$R_{\rm{\gamma}}+R_{\rm{BLR}}-R_{\rm{\gamma}}c/v_{\rm{d}}=c\tau_{\rm{ob}}/(1+z)$,
and then
\begin{equation}
R_{\rm{\gamma}}=\frac{R_{\rm{BLR}}- \frac{c
\tau_{\rm{ob}}}{1+z}}{\frac{c}{v_{\rm{d}}}-1},
\end{equation}
where $z$ is the redshift of source, $\tau_{\rm{ob}}>0$ and
$\tau_{\rm{ob}}$ is the observed time lag of the broad lines
relative to the $\gamma$-rays.

Outside segment $AG$, the $\gamma$-rays will lag the lines. As the
disturbances reach point $H$, where the $\gamma$-rays are
produced, i.e. $R_{\rm{\gamma}}=AH$, the line photons reach point
$L$. The light travelling time effects for the $\gamma$-ray
photons from point $J$ to $L$, i.e. from point $H$ to $M$, result
in the time lag of the $\gamma$-rays relative to the lines (see
Fig. 1$a$). In this case (hereafter Case C), we have
$R_{\rm{\gamma}}c/v_{\rm{d}}=R_{\rm{\gamma}}+R_{\rm{BLR}}+c\tau_{\rm{ob}}/(1+z)$,
and then
\begin{equation}
R_{\rm{\gamma}}=\frac{R_{\rm{BLR}}+\frac{c
\tau_{\rm{ob}}}{1+z}}{\frac{c}{v_{\rm{d}}}-1},
\end{equation}
where $\tau_{\rm{ob}}>0$ and $\tau_{\rm{ob}}$ is the observed time
lag of the $\gamma$-rays relative to the broad lines. Equations
(1), (2) and (3) can be unified into
\begin{equation}
R_{\rm{\gamma}}=\frac{R_{\rm{BLR}}+\frac{c
\tau_{\rm{ob}}}{1+z}}{\frac{c}{v_{\rm{d}}}-1},
\end{equation}
where $\tau_{\rm{ob}}$ is the observed time lags of the
$\gamma$-rays relative to the broad lines and is zero, negative or
positive. As $\tau_{\rm{ob}}=0$ (Case A), equation (4) becomes
equation (1). As $\tau_{\rm{ob}}<0$ (Case B), equation (4) becomes
equation (2). As $\tau_{\rm{ob}}>0$ (Case C), equation (4) becomes
equation (3). Once $\tau_{\rm{ob}}$, $R_{\rm{BLR}}$ and
$v_{\rm{d}}$ are known, $R_{\rm{\gamma}}$ can be obtained from
equations (1)--(4).

The calculations above are under the condition of $\theta = 0$,
where $\theta$ is the angle between the jet axis and the line of
sight. In fact, the approaching relativistic jets of blazars are
oriented at a small angle to the line of sight \citep{b14}. Thus
$\theta \neq 0$. Considering the actual inclination of the jet
axis with respect to the line of sight (see Fig. 1$b$), we
re-deduce the expression of Case C. As the disturbances reach
point $H$, where the $\gamma$-rays are produced, the line photons
reach point $L$. The light travelling time effects for the
$\gamma$-ray photons from point $O$ to $L$, i.e. from point $H$ to
$S$, result in the time lag of the $\gamma$-rays relative to the
lines (see Fig. 1$b$). In this case, we have
$R_{\rm{\gamma}}c/v_{\rm{d}}=R_{\rm{BLR}}+R_{\rm{\gamma}}\cos
\theta +c\tau_{\rm{ob}}/(1+z)$, and then
\begin{equation}
R_{\rm{\gamma}}=\frac{R_{\rm{BLR}}+\frac{c
\tau_{\rm{ob}}}{1+z}}{\frac{c}{v_{\rm{d}}}-\cos \theta}.
\end{equation}
It is obvious that equation (5) contains Cases A, B and C. As
$\theta = 0$, equation (5) becomes equation (4). The observed line
photons are from the BLR of ring, and then the observed lag is an
ensemble average over all points of the ring. For point $T$, we
have
$R_{\rm{\gamma}}c/v_{\rm{d}}=R_{\rm{BLR}}+TW+R_{\rm{\gamma}}\cos
\theta +c\tau_{\rm{ob}}/(1+z)$ and $TW=R_{\rm{BLR}} \sin \alpha
\sin \theta$ (see Fig. 1$b$), and then
\begin{equation}
R_{\rm{\gamma}}=\frac{R_{\rm{BLR}}(1+\sin \alpha \sin
\theta)+\frac{c \tau_{\rm{ob}}}{1+z}}{\frac{c}{v_{\rm{d}}}-\cos
\theta},
\end{equation}
where $\alpha$ is the angle between $AB$ and $AT$ and varies from
0 to $2\pi$. For a given source, i.e. given $R_{\rm{\gamma}}$,
$R_{\rm{BLR}}$, $v_{\rm{d}}$, $\theta$ and $z$, $\tau_{\rm{ob}}$
varies with $\alpha$, i.e.
$\tau_{\rm{ob}}=\tau_{\rm{ob}}(\alpha)$. Calculating ensemble
average over $\alpha$ in equation (6), we have
\begin{equation}
R_{\rm{\gamma}}=\frac{R_{\rm{BLR}}+\frac{c \langle
\tau_{\rm{ob}}\rangle}{1+z}}{\frac{c}{v_{\rm{d}}}-\cos \theta},
\end{equation}
where $\langle \tau_{\rm{ob}} \rangle$ is the ensemble average of
$\tau_{\rm{ob}}(\alpha)$ and is the measured time lag between the
$\gamma$-ray and line light curves (and $\langle 1+\sin \alpha
\sin \theta \rangle=\int_{0}^{2\pi}(1+\sin \alpha \sin \theta)
\rm{d} \alpha /2\pi=1$). It is obvious that equation (7) contains
Cases A, B and C. As $\theta =0$, equation (7) becomes equation
(4). In the following sections, we will calculate
$R_{\rm{\gamma}}$ based on equation (7).

\begin{figure}
\begin{centering}
\includegraphics[width=2.0 in,angle=-90]{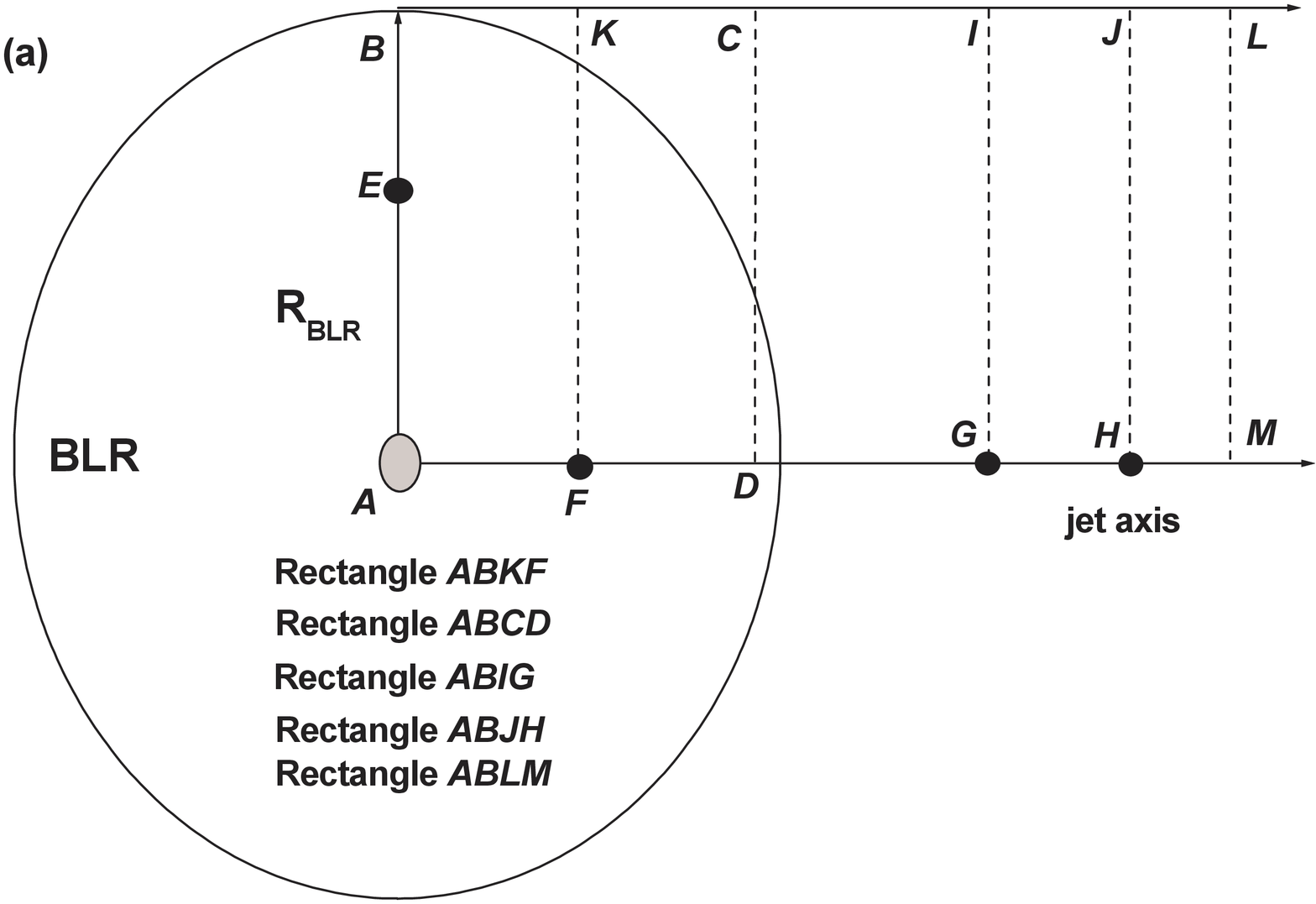}
\includegraphics[width=2.0 in,angle=-90]{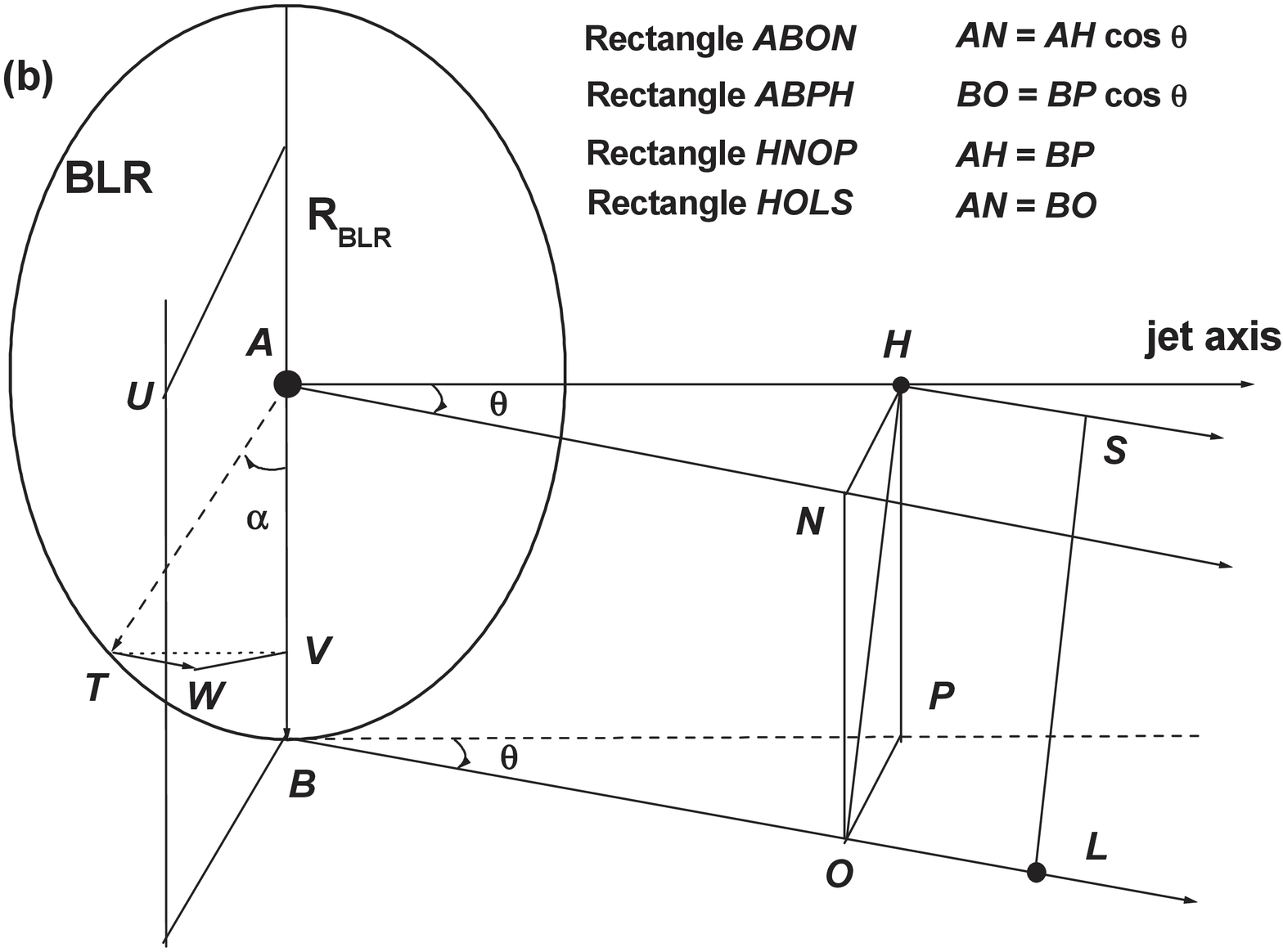}
\end{centering}
 \caption{Sketch of the geometry assumed, and it is similar to that used in the reverberation mapping
 method of broad emission lines. $R_{\rm{BLR}}$ is the size of BLR. ($a$) the angle between the line of
 sight and the jet axis $\theta=0$. ($b$) $\theta \neq 0$. The jet axis is perpendicular to the plane
 of BLR. ($b$) the planes $ABU$ and $HNOP$ are perpendicular to the
 line of sight. $\because$ $TV\perp AV$, $\therefore$ $TV=AT\sin \alpha=R_{\rm{BLR}}\sin \alpha$.
 $\because$ $TW\parallel BL$, $\therefore$ $TW \perp WV$ and $TW \perp AV$. $TW \perp AV$ and $TV \perp AV$
 give $WV \perp AV$. $\because$ $TV \perp AV$ and $WV \perp AV$, $\therefore$ $\angle TVW=\theta$. $\because$
 $\angle TVW=\theta$ and $\angle TWV=\pi /2$, $\therefore$
 $TW=TV\sin \theta =R_{\rm{BLR}}\sin \alpha \sin \theta$. }
  \label{fig1}
\end{figure}

\section{APPLICATION TO 3C 273}
3C 273 was first identified as a quasar at redshift $z=0.158$ by
\citet{b74}. It is one of the best studied AGNs in all bands
\citep[see e.g.][]{b53,b90,b85,b80}. The database site\footnote
{http://isdc.unige.ch/3c273/} of 3C 273 provides a series of about
70 light curves as well as some spectra. The jet of 3C 273 is
one-sided, with no signs of emission from the counterjet side
\citep{b88}. The blue-bump of 3C 273 is thermal continuum emission
from the inner accretion disc \citep{b75}. Fe K$\alpha$ lines
observed in 3C 273 were shown to be from an accretion disc around
a supermassive black hole \citep{b95,b84}. If the assumptions in
the method are correct, it is expected for 3C 273 that there
should exist time lags between the broad lines from the BLR and
the $\gamma$-rays from the relativistic jet.

\subsection{Data of 3C 273}

This paper makes use of the 3C 273 database hosted by the ISDC
\citep{b85}. This database is one of the most complete
multi-wavelength databases currently available for one AGN. For 3C
273, the $\gamma$-ray light curves are very sparsely sampled
and/or the error bars of $\gamma$-ray fluxes are also very large.
Of course with {\it Fermi\/} taking data since June 2008 this is
no longer the case \citep[see e.g.][]{b3}. However, the sampling
over the dates of interest, and, in general over the timescales of
interest is still sparse (see the line light curves in the
following paragraphs). Thus it should be unreliable to employ the
$\gamma$-ray light curves to estimate the time lags. Hence for
these blazars lacking adequate $\gamma$-ray light curves, the
synchrotron emission, especially the radio emission, could be used
to derive the time lags relative to the broad lines. The
synchrotron flares from the relativistic jet dominate energy
output from radio to millimeter and extend up to the
infrared--optical regimes \citep{b70,b86,b80}. Radio light curves
are better than millimeter ones in the samplings and the features
of synchrotron flares. Gamma-ray detections correspond to rising
radio fluxes \citep[e.g.][]{b87}. Thus the radio light curves are
adopted to be analyzed. Though the synchrotron flares extend up to
the infrared--optical regimes, the flares are sparse (denoted by
red color in the 3C 273 database). In the intervals among these
flares, the infrared--optical light curves may be contaminated by
other emission components. Thus it is expected that there should
be no significant features in the cross correlation functions
between the broad lines and the light curves. Considering that the
synchrotron emission peaks around the infrared band for 3C 273
\citep[see e.g.][]{b34}, the infrared light curves are also
adopted to be analyzed and to test the expectation above. In the
light curves adopted here, only good data (Flag$>=$0) are adopted.

Light curves of 5, 8, 15, 22 and 37 GHz are taken from the 3C 273
database, for these light curves have enough data \citep{b85}.
These radio light curves are presented in Fig. 2. There are four
distinct outbursts in each radio light curve after $\sim$ 1980
(see Fig. 2). The data of 22 and 37 GHz are adopted from $\sim$
1980, for observations are very sharply sampled before $\sim$ 1980
(see Fig. 2). From 1973 to 1978, there is a gap of 5 years without
observations of 5 GHz. Considering the four distinct outbursts in
each radio light curve, the data from $\sim$ 1980 are adopted for
the light curve of 5 GHz. Also, the data from $\sim$ 1980 are
adopted for the light curves of 8 and 15 GHz. The sampling rates
of 5, 8, 15, 22 and 37 GHz are 29, 40, 40, 44 and 46 times per
year for the adopted data, respectively.

Light curves of broad lines H$\alpha$, H$\beta$ and H$\gamma$ are
taken from \cite{b46}, and all the data in the light curves are
adopted to calculate the time lags relative to the radio emission.
The sampling rates of the lines are around 5 times per year. The
line light curves are also presented in Fig. 2, wherein all the
light curves show the same time interval. It can be seen in Fig. 2
that the line light curves are sharply sampled relative to the
radio light curves adopted. Especially, there is only one data
point in each of several valley-bottoms in the H$\alpha$ and
H$\beta$ light curves, and the trend of variations can be changed
if the points in these valley-bottoms are excluded from the
H$\alpha$ and H$\beta$ light curves (see Fig. 2). The sharp
samplings will affect the choice of analysis method for the cross
correlation between the broad lines and the radio emission. The
infrared light curves of J, H, K and L bands are also adopted from
the 3C 273 database. The data before 1998 are adopted for the J, H
and L light curves. The data before 2000 are adopted for the K
light curve.
\begin{figure*}
\begin{center}
\includegraphics[width=4.0 in,angle=-90]{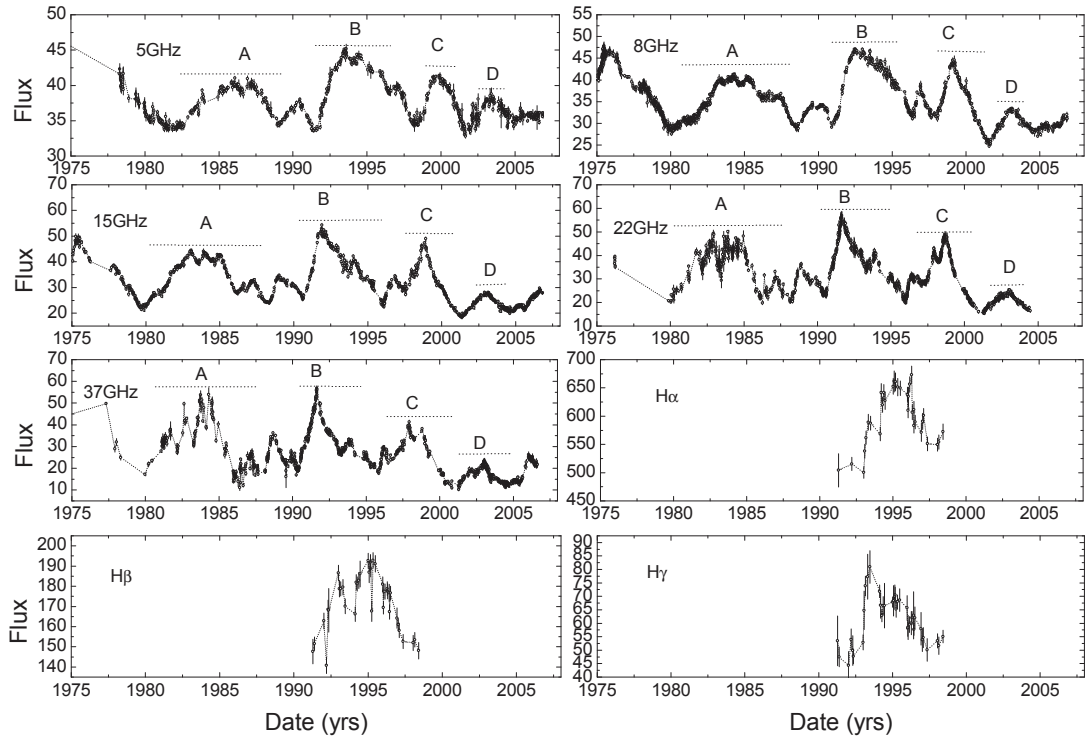}
 \end{center}
 \caption{Light curves of the radio emission and the Balmer lines. The y-axis is in units of
 Jy. For lines, the y-axis is in units of
 $10^{-14}\/\ \rm{erg\/\ cm^{-2}}\/\ s^{-1}$. A, B, C and D denote the four distinct
 outbursts after $\sim$ 1980.
 }
  \label{fig2}
\end{figure*}

\subsection{Analysis of time lags}
Cross-correlation function (CCF) analysis is a standard technique
in time series analysis for finding time lags between light curves
at different wavelengths, and the definition of the CCF assumes
that the light curves are uniformly sampled. However, in most
cases the sampling is not uniform. The interpolated cross
correlation function (ICCF) method of \citet{b31} uses a linear
interpolation scheme to determine the missing data in the light
curves. On the other hand, the discrete correlation function
\citep[DCF;][]{b25} can utilize a binning scheme to approximate
the missing data. Apart from the ICCF and DCF, there is another
method of estimating the CCF in the case of nonuniformly sampled
light curves, the z-transformed discrete correlation function
\citep[ZDCF;][]{b6}. The ZDCF is a binning type of method as an
improvement of the DCF technique, but it has a notable feature in
that the data are binned by equal population rather than equal bin
width as in the DCF. It has been shown in practice that the ZDCF
is more robust than the ICCF and the DCF when applied to sparsely
and unequally sampled light curves \citep[see
e.g.][]{b26,b39,b71}. \citet{b57} analyzed the ZDCFs between
unequally sampled light curves of AGNs, and they obtained
inter-band time lags. In practice, the ZDCF is applicable and
reliable to analyze the unequally sampled light curves. Thus the
ZDCF will be calculated in the paper because of the sharp
samplings of the Balmer lines.

In general, it seems to be true that the time lag is better
characterized by the centroid $\tau_{\rm{cent}}$ of the DCF and
the ICCF than by the peak value $\tau_{\rm{peak}}$, namely, the
time lag where the linear correlation coefficient has its maximum
value $r_{\rm{max}}$ \citep[see e.g.][]{b65,b66}. In both the DCF
and the ICCF, $\tau_{\rm{peak}}$ is much less stable than
$\tau_{\rm{cent}}$, but $\tau_{\rm{peak}}$ is much less stable in
the DCF than in the ICCF \citep{b66}. Thus we prefer the time lag
to be characterized by the centroid $\tau_{\rm{cent}}$ of the
ZDCF. The centroid time lag $\tau_{\rm{cent}}$ is computed using
all the points with correlation coefficients $r \geqslant 0.8
r_{\rm{max}}$ in the ZDCF bumps closer to the zero-lag (see Fig.
3). The calculated ZDCFs between the radio and broad-line light
curves are presented in Fig. 3. The horizontal and vertical error
bars in Fig. 3 represent the 68.3\% confidence intervals in the
time lags and the relevant correlation coefficients, respectively.
The ZDCF bumps closer to the zero-lag have a good profile in Fig.
3. The measured time lags are listed in Table 1. The centroid
$\tau_{\rm{cent}}$ is calculated by
$\tau_{\rm{cent}}=\sum\tau(i)r(i)/\sum r(i)$, where $\tau(i)$ and
$r(i)$ are the values of $i$-th data pair with $r \geqslant 0.8
r_{\rm{max}}$. The errors of $\tau_{\rm{cent}}$ are calculated by
$\Delta \tau^{\pm}_{\rm{cent}}=\{\sum[\Delta \tau^{\pm}(i)
r(i)+\tau(i)\Delta r^{\pm}(i)]\sum r(i)-\sum \tau(i)r(i) \sum
\Delta r^{\pm}(i)\}/[\sum r(i)]^2$, where $\Delta \tau^{\pm}(i)$
and $\Delta r^{\pm}(i)$ are the relevant errors of $\tau(i)$ and
$r(i)$, respectively.
\begin{table*}
\centering
 \begin{minipage}{100mm}
  \caption{Time lags between emission lines and
radio emission. The sign of the time lag is defined as
$\tau_{\rm{cent}}= t_{\rm{radio}}-t_{\rm{line}}$. Time lags are in
units of yrs.}
  \begin{tabular}{lrrrrr}

   \hline \hline

 Lines& 5 GHz & 8 GHz&15 GHz &22 GHz&37 GHz\\

 \hline
H$\alpha$&$-1.23^{+0.02}_{-0.07}$&$-1.99^{+0.06}_{-0.01}$&$-2.77^{+0.02}_{-0.08}$&$-3.14^{+0.02}_{-0.09}$&$-3.30^{+0.02}_{-0.08}$\\
H$\beta$&$-0.19^{+0.03}_{-0.09}$&$-1.06^{+0.08}_{-0.02}$&$-2.40^{+0.02}_{-0.07}$&$-2.27^{+0.03}_{-0.09}$&$-3.27^{+0.02}_{-0.08}$\\
H$\gamma$&$-0.35^{+0.03}_{-0.09}$&$-1.13^{+0.07}_{-0.01}$&$-1.57^{+0.02}_{-0.07}$&$-1.82^{+0.02}_{-0.08}$&$-2.01^{+0.02}_{-0.07}$\\

H$\alpha$&$4.72^{+0.08}_{-0.02}$&$4.20^{+0.06}_{-0.02}$&$3.43^{+0.08}_{-0.02}$&$3.21^{+0.07}_{-0.02}$&$3.06^{+0.08}_{-0.03}$\\
H$\beta$&$4.31^{+0.07}_{-0.02}$&$4.13^{+0.07}_{-0.01}$&$3.72^{+0.07}_{-0.02}$&$3.45^{+0.07}_{-0.02}$&$3.26^{+0.09}_{-0.03}$\\
H$\gamma$&$6.19^{+0.10}_{-0.04}$&$4.13^{+0.06}_{-0.01}$&$3.73^{+0.07}_{-0.02}$&$3.78^{+0.08}_{-0.03}$&$3.29^{+0.08}_{-0.03}$\\

\hline

\end{tabular}
\end{minipage}
\end{table*}

Our results show two possibilities that the broad-line variations
lag or lead the radio ones (see Fig. 3). The measured time lags
are on the order of years (see Table 1). For a given line, the
relevant time lags generally decrease as radio frequency increases
from 5 to 37 GHz (see Table 1). The calculated ZDCFs between the
infrared and H$\alpha$ light curves are presented in Fig. 4 for
illustration. For the ZDCFs in Fig. 4, there are no significant
common features closer to the zero-lag. Also, there are no
significant common features closer to the zero-lag for the ZDCFs
of the H$\beta$ and H$\gamma$ lines relative to the infrared
emission. On the contrary, the ZDCFs in Fig. 3 have significant
common features closer to the zero-lag. The absence of significant
features tests the expectation in subsection 3.1. This test
indicates that the infrared synchrotron emission does not dominate
the energy output in the intervals between the synchrotron flares
in the infrared--optical bands. The lag $\tau_{\rm{ob}}$ is the
lag $\tau_{\rm{cent}}$ measured here. Hereafter,
$\tau_{\rm{cent}}$ is equivalent to $\tau_{\rm{ob}}$ and $\langle
\tau_{\rm{ob}} \rangle$.
\begin{figure*}
 \begin{center}
\includegraphics[width=4.0 in,angle=-90]{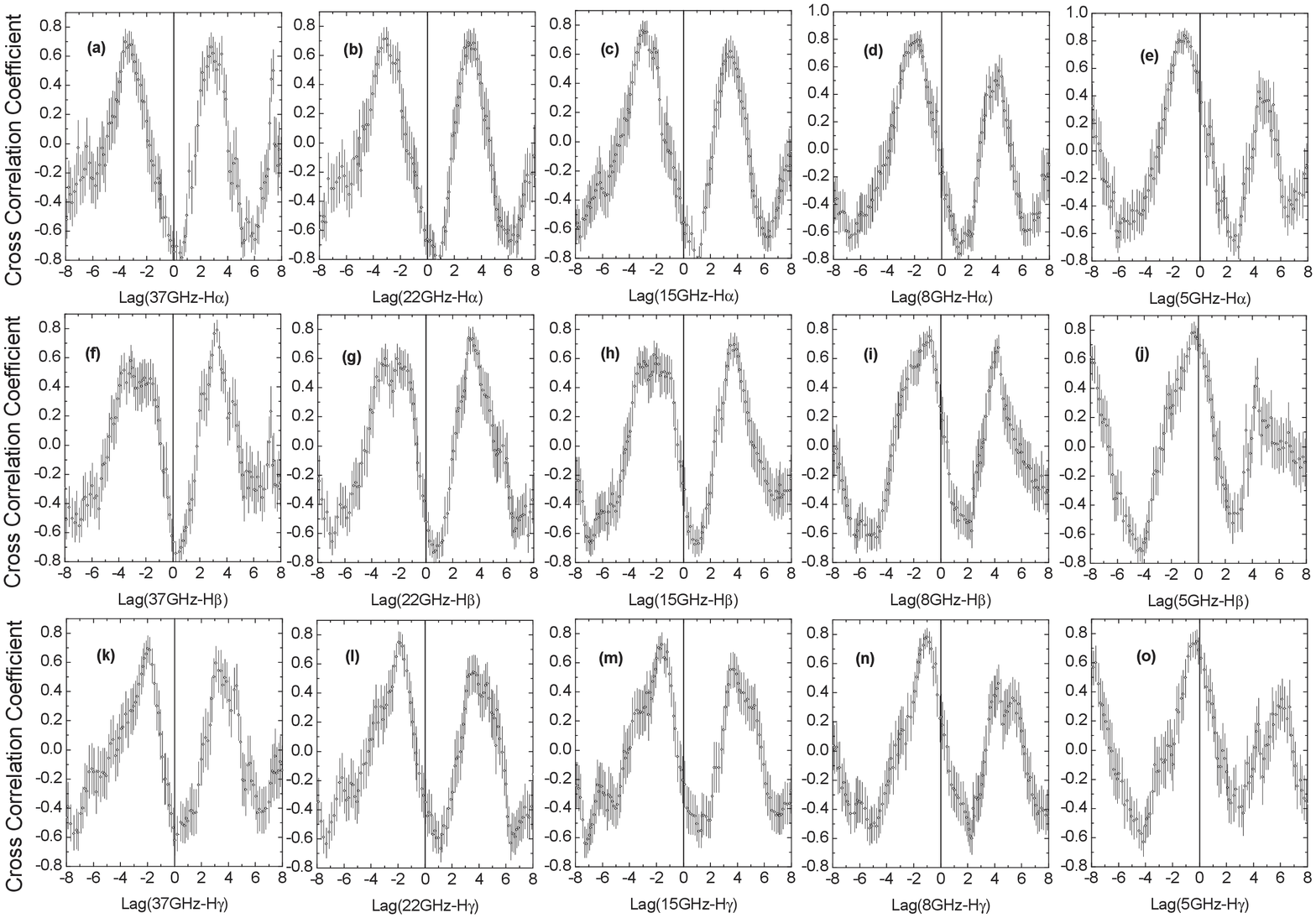}
 \end{center}
 \caption{ZDCF between H$\alpha$ and $(a)$ 37, $(b)$ 22, $(c)$ 15, $(d)$ 8 and $(e)$ 5 GHz;
 ZDCF between H$\beta$ and $(f)$ 37, $(g)$ 22, $(h)$ 15, $(i)$ 8 and $(j)$ 5 GHz;
 ZDCF between H$\gamma$ and $(k)$ 37, $(l)$ 22, $(m)$ 15, $(n)$ 8 and $(o)$ 5
 GHz. The x-axis is in units of yrs.
 }
  \label{fig3}
\end{figure*}
\begin{figure*}
\begin{center}
\includegraphics[width=4.0 in,angle=-90]{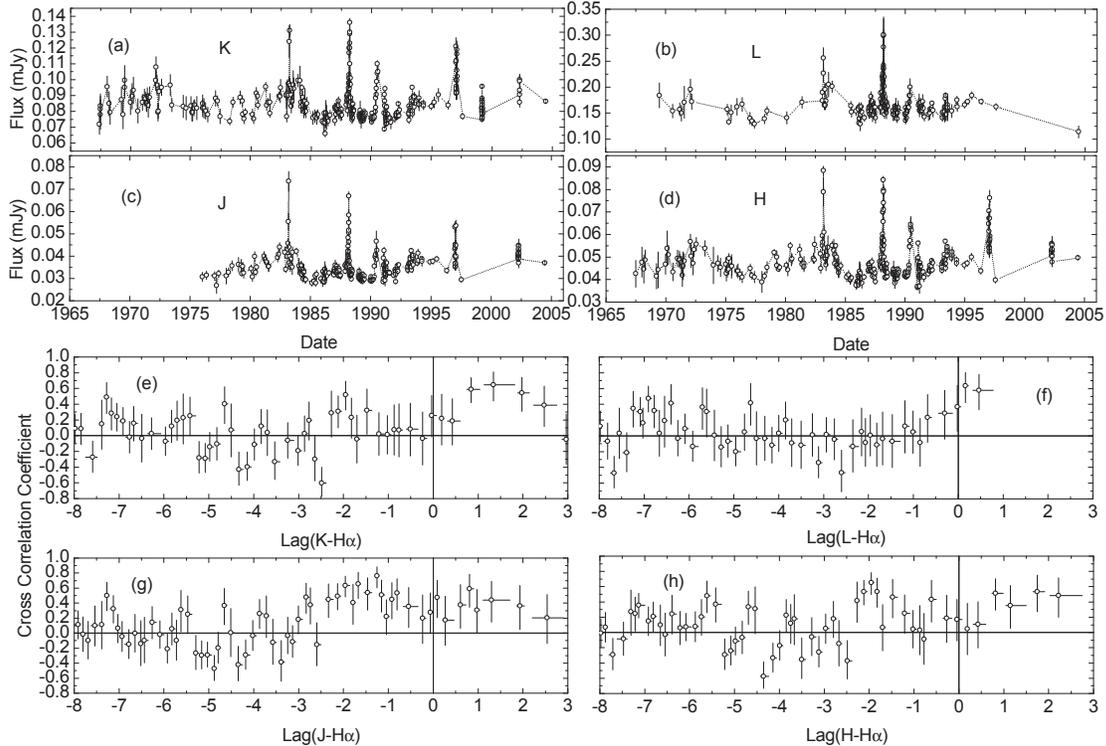}
 \end{center}
 \caption{Light curves of infrared emission in $(a)$ K, $(b)$ L, $(c)$ J and $(d)$ H
 bands. ZDCF between H$\alpha$ and $(e)$ K, $(f)$ L, $(g)$ J and $(h)$ H
 bands. The x-axis is in units of yrs.
 }
  \label{fig4}
\end{figure*}

\subsection{Calculations}

For 3C 273, \citet{b46} determined the H$\alpha$, H$\beta$ and
H$\gamma$ lags $\tau$ relative to the optical continuum, and
\citet{b62} determined their lags relative to the UV continuum.
The optical continuum is strongly contaminated by non-thermal
emission, possibly related to the relativistic jet, and therefore
it appears unsuitable for studying the lags between the ionizing
continuum and the lines \citep{b61}. The H$\alpha$, H$\beta$ and
H$\gamma$ lags relative to the UV continuum are more reliable than
those relative to the optical continuum \citep{b62}. Thus we adopt
the H$\alpha$, H$\beta$ and H$\gamma$ lags determined by
\citet{b62}. Here, the average of rest-frame lags of these lines,
$\overline \tau=2.70$ years, is adopted as a characteristic value
of $\tau$. Thus the BLR has a typical size of $R_{\rm{BLR}}=2.70$
$\rm{ly}$.

Simulations show that the relativistic jets can be driven from a
region just outside the ergosphere of a Kerr black hole \citep[see
e.g.][]{b60,b47,b48,b49,b72}. In most cases, the bulk velocity of
jet $v_{\rm{j}}$ is close to the escape speed \citep{b51}. The
escape speed is around $0.9c$ near the ergosphere of the rapidly
spinning black hole \citep{b60}. Most supermassive black holes are
spinning rapidly \citep{b27}. Thus $v_{\rm{j}}\sim 0.9c$. If the
disturbances in the cental engine are transported with the jet
itself, $v_{\rm{d}}=v_{\rm{j}}\sim 0.9c$. Thus we would have
$R_{\rm{\gamma}}\sim 9 R_{\rm{BLR}}$ from equation (1).

For 3C 273, the jet has $v_{\rm{j}}= 0.95c$ and $\cos \theta
=0.95$ on 100 pc scales \citep{b24a}. The pc-scale jet was
constrained to have $\theta <15^{\circ}$ and the bulk Lorentz
factor $\Gamma > 10$ \citep{b88}. The actual value of $\theta$
cannot be too small, unlike better aligned blazars, because it has
a strong big blue bump \citep{b75,b23}. Thus it is likely that
$\theta$ be larger than $10^{\rm{\circ}}$. The ratio of the jet to
counterjet flux is $R=[(1+v_{\rm{j}}\cos \theta
/c)/(1-v_{\rm{j}}\cos \theta /c)]^{\rm{3+\alpha}}$ for discrete
moving blobs \citep{b54a}. $R>10^4$ was observed for 3C 273 and
$v_{\rm{j}}$ can be up to $0.995c$ \citep[see e.g.][]{b32a}. The
observed spectral index $\alpha =0.8$ \citep{b88}. It is obvious
that $\theta \la 21^{\circ}$ and $0.9c\leq v_{\rm{j}}\leq 0.995c$
are allowed by $R>10^4$. The disturbances are transported from the
central engine down the jet, and then it is possible that
$v_{\rm{d}}=v_{\rm{j}}=0.9$--$0.995c$ and
$\theta=12^{\rm{\circ}}$--$21^{\rm{\circ}}$. For a given line, the
relevant time lags generally decrease as radio frequency increases
from 5 to 37 GHz (see Table 1). We adopt the measured time lags of
the lines relative to the 37 GHz emission. The negative lags have
an average of $\overline \tau^{-}_{\rm{ob}}=-2.86$ years. The
positive lags have an average of $\overline
\tau^{+}_{\rm{ob}}=3.20$ years. There is no zero-lag. In the
following calculations, $v_{\rm{d}}=0.9$--$0.995c$ and
$\theta=12^{\rm{\circ}}$--$21^{\rm{\circ}}$ are adopted, and these
values estimated from $\tau_{\rm{ob}}$ are denoted by
$R_{\rm{radio}}$.

From $\overline \tau^{-}_{\rm{ob}}=-2.86$ years,
$R_{\rm{BLR}}=2.70$ $\rm{ly}$, $v_{\rm{d}}=0.9$--$0.995c$,
$\theta=12^{\rm{\circ}}$--$21^{\rm{\circ}}$ and equation (7), we
can obtain the radio emitting position $R_{\rm{radio}}=0.40$--2.62
pc (Case B). The typical size of $R_{\rm{BLR}}=2.70$
$\rm{ly}=0.83$ pc is within the range of $R_{\rm{radio}}$
estimated in Case B. The radio emitting regions in Case B are at
distances of pc-scale from the central engine and are around the
BLR, i.e. likely inside the BLR, co-located with the BLR or
outside the BLR. From $\overline \tau^{+}_{\rm{ob}}=3.20$ years,
$R_{\rm{BLR}}=2.70$ $\rm{ly}$, $v_{\rm{d}}=0.9$--$0.995c$,
$\theta=12^{\rm{\circ}}$--$21^{\rm{\circ}}$ and equation (7), we
can obtain $R_{\rm{radio}}=9.43$--62.31 pc (Case C). The estimated
sizes are much larger than the typical size of $R_{\rm{BLR}}=0.83$
pc. The radio emitting regions in Case C are at distances of tens
of pc from the central black hole and are far away from the BLR.

\citet{b50} identified the pc-scale radio core as a likely
location for both the $\gamma$-ray and radio flares. \citet{b41}
concluded that both the radio and $\gamma$-ray events are
originating from the same region of a relativistic jet. In 1990s,
it is commonly thought that the $\gamma$-rays are produced in the
jet, but closer to the central engine than the radio emission
\citep[see e.g.][]{b24}. Thus it is expected that
$R_{\rm{\gamma}}\la R_{\rm{radio}}$. The constraint of
$R_{\rm{\gamma}}\la R_{\rm{radio}}$ is allowed by the recent
flares of 3C 279 observed by {\it Fermi} and in a multi-wavelength
campaign \citep{b5}, where radio light curves from 5 to 230 GHz
fail to show prominent variations during either the November 2008
or the February 2009 $\gamma$-ray flares (or anytime in between).
For Case B, $R_{\rm{\gamma}}\la 0.40$--2.62 pc for
$v_{\rm{d}}=0.9$--$0.995c$ and
$\theta=12^{\rm{\circ}}$--$21^{\rm{\circ}}$. For Case A, the
zero-lag position $R_{\rm{\gamma}}=4.67$--30.81 pc, which is far
away from the BLR. For Case C, $R_{\rm{\gamma}}\la 9.43$--62.31
pc. Also, $R_{\rm{\gamma}}>4.67$--30.81 pc for the positive lags
in Case C. Thus for Case C we have 4.67--$30.81<R_{\rm{\gamma}}\la
9.43$--62.31 pc for $v_{\rm{d}}=0.9$--$0.995c$ and
$\theta=12^{\rm{\circ}}$--$21^{\rm{\circ}}$. The dependence of
$R_{\rm{\gamma}}$ and $R_{\rm{radio}}$ on $v_{\rm{d}}$ and
$\theta$ is presented in Fig. 5 (see three-dimensional plots).
$R_{\rm{\gamma}}$ and $R_{\rm{radio}}$ increase as $v_{\rm{d}}$
increases, but decrease as $\theta$ increases. The uncertainties
of $v_{\rm{d}}$ and $\theta$ result in the larger intervals of
$R_{\rm{\gamma}}$ and $R_{\rm{radio}}$. For better representing
intervals of $R_{\rm{\gamma}}$, the sections of the
three-dimensional plots at $\cos \theta =0.95$ are also plotted in
Fig. 5 (see the bottom panel). For Case B, $R_{\rm{\gamma}}\la
R_{\rm{radio}}=0.44$--1.28 pc for $v_{\rm{d}}=0.9$--$0.995c$ and
$\cos \theta=0.95$. These $R_{\rm{\gamma}}$ marginally satisfy
$R_{\rm{\gamma}}\la R_{\rm{BLR}}$. For Case C, we have
5.15--$15.08<R_{\rm{\gamma}}\la 10.40$--30.45 pc.
\begin{figure}
\begin{centering}
\includegraphics[width=2.0 in,angle=-90]{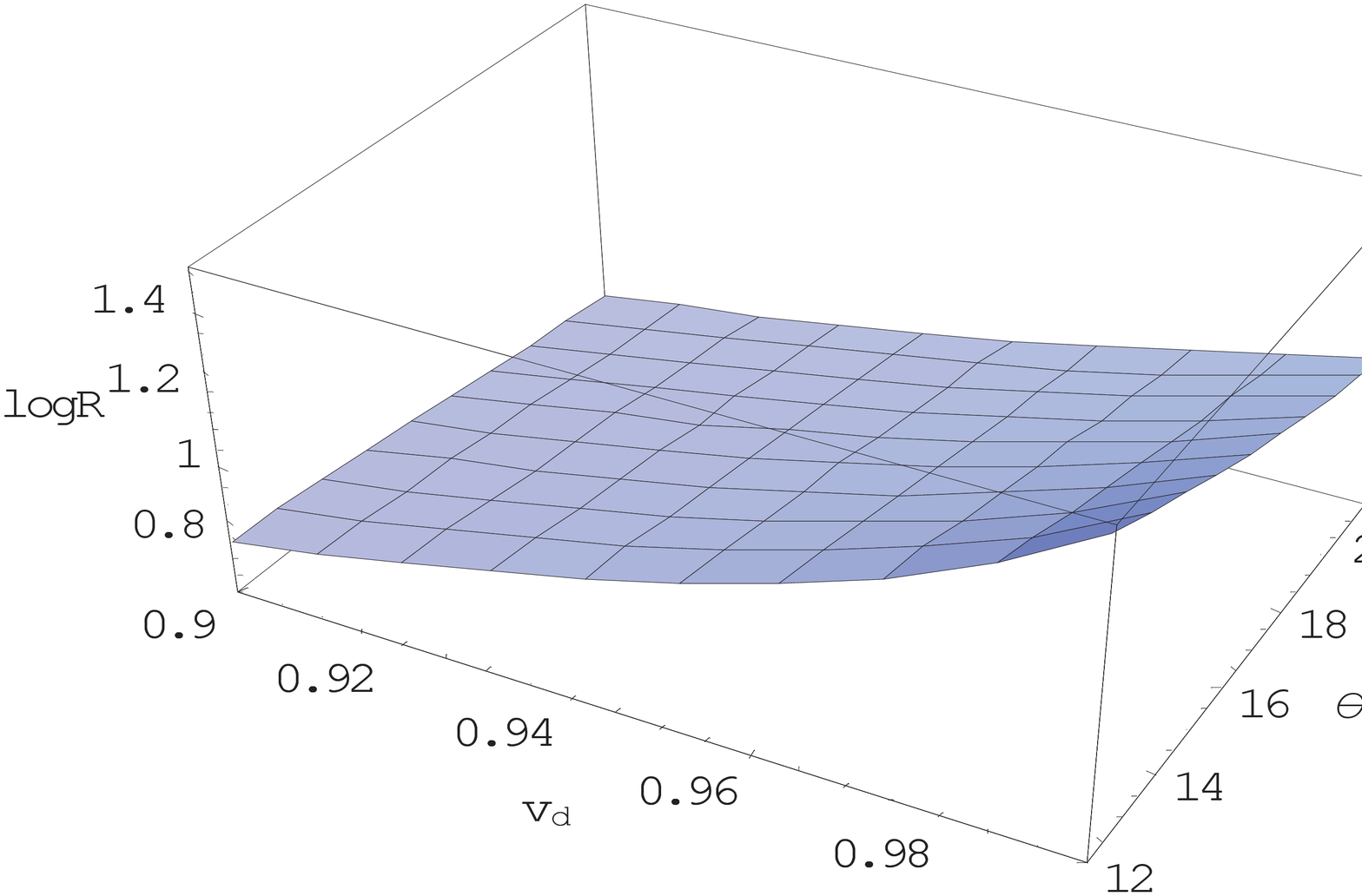}
\includegraphics[width=2.0 in,angle=-90]{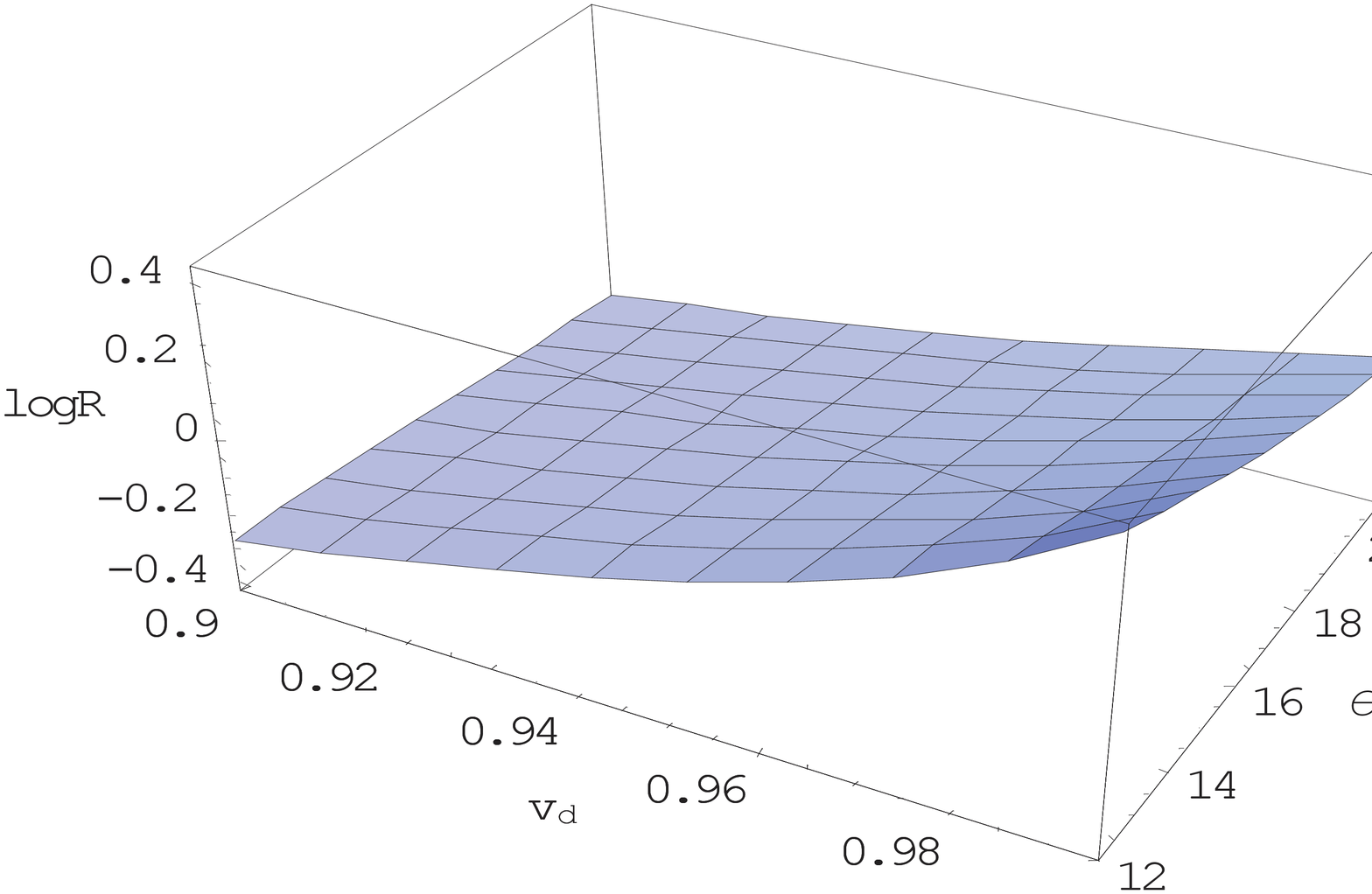}
\includegraphics[width=2.0 in,angle=-90]{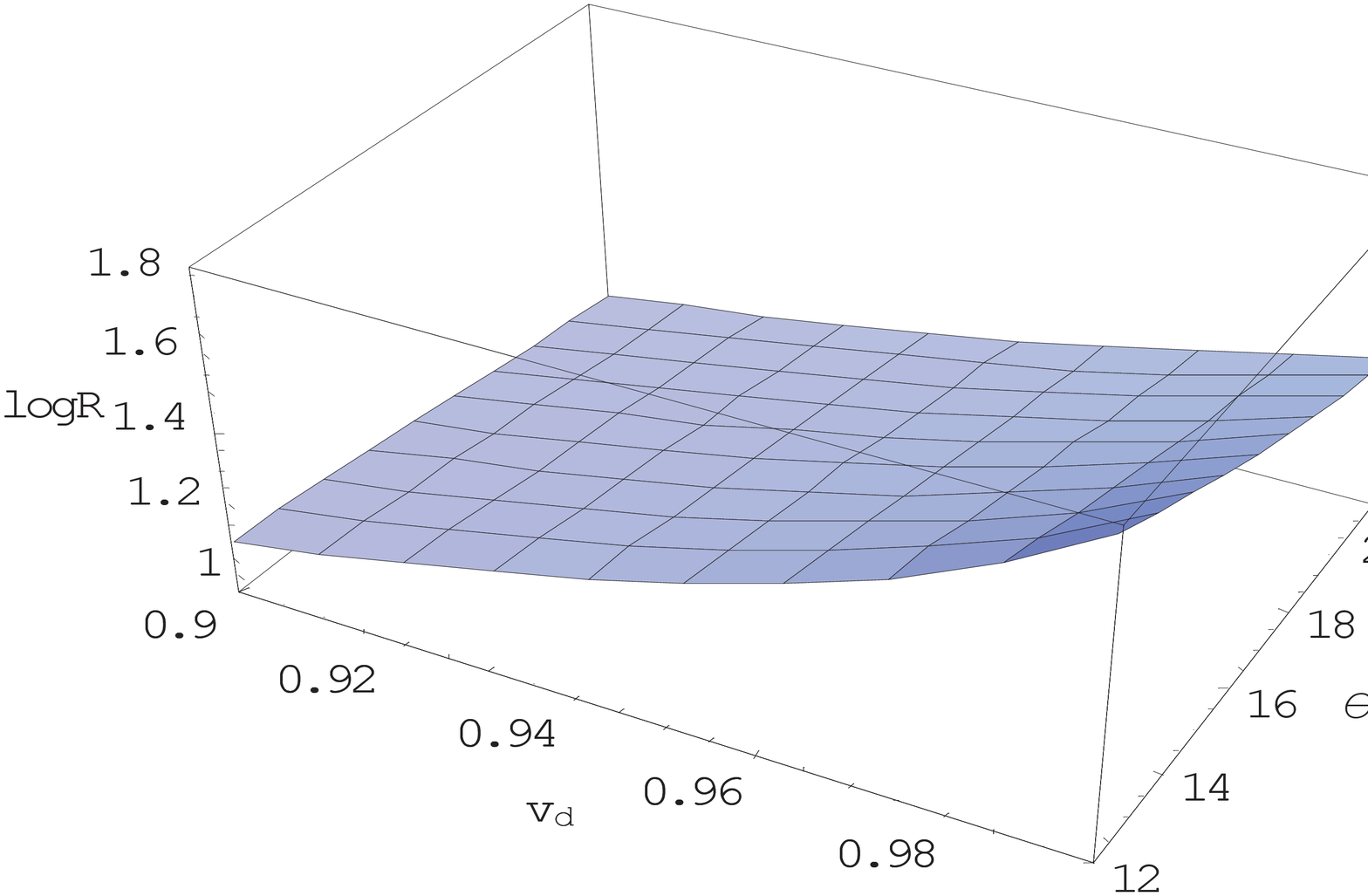}
\includegraphics[width=2.0 in,angle=-90]{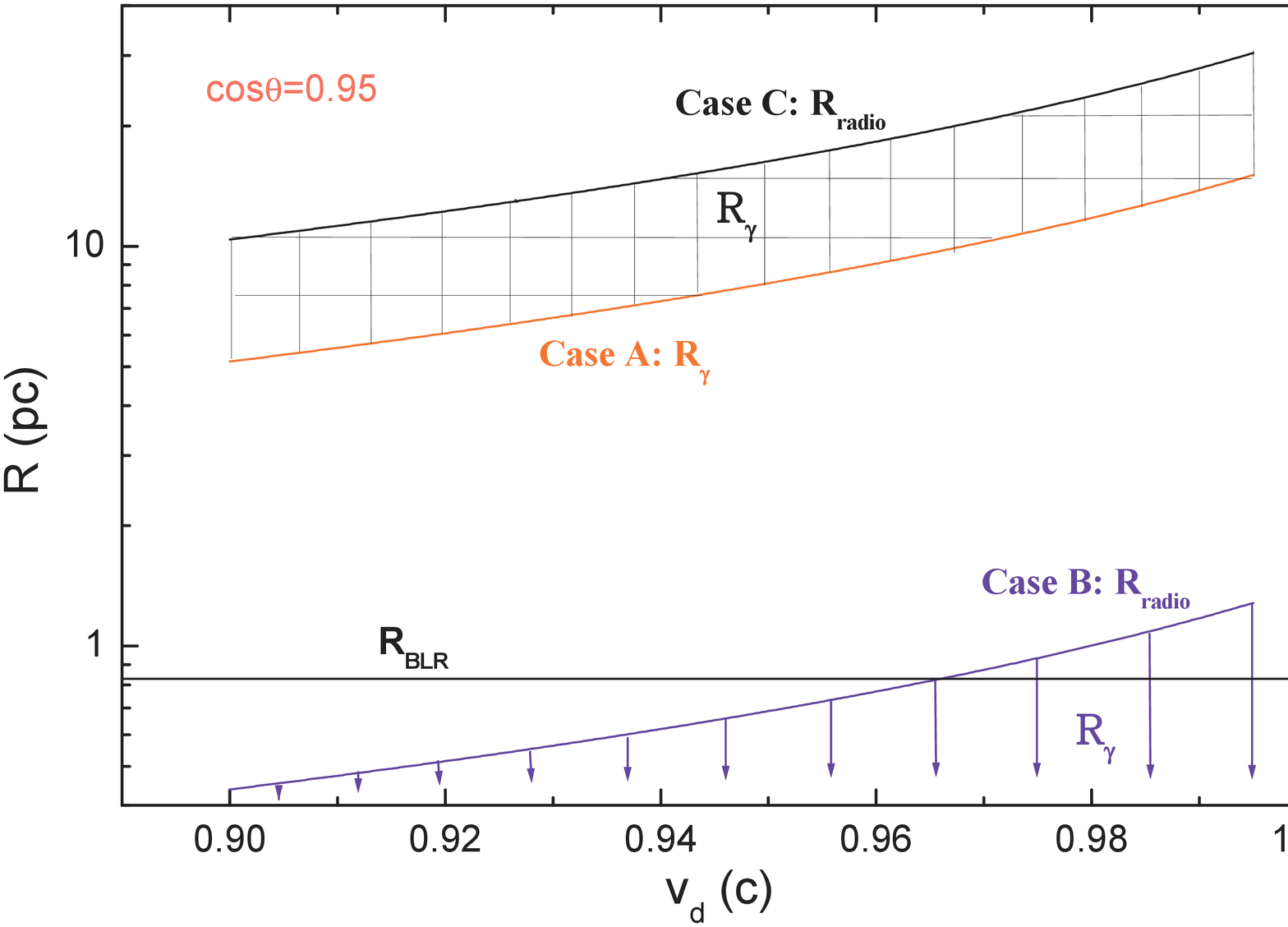}
\end{centering}
\caption{Dependence of distance $R$ on $v_{\rm{d}}$ and $\theta$.
From the top down, the first three panels correspond to Cases A, B
and C, respectively. The $x$, $y$ and $z$-axes are $v_{\rm{d}}$ in
units of $c$, $\theta$ in units of degree and $\log R$ in units of
pc, respectively. The first three panels represent dependence of
$R$ on $v_{\rm{d}}$ and $\theta$. The bottom panel represents
dependence of $R$ on $v_{\rm{d}}$ in case of $\cos \theta =0.95$,
where the gridding area represents the allowed interval of
$R_{\rm{\gamma}}$ in Case C. The arrows represent the upper limit
of $R_{\rm{\gamma}}$ in Case B.
 }
  \label{fig5}
\end{figure}

It is possible that there is a special point $D$ within segment
$AG$ (see Fig. 1$a$). As the ionizing photons travel from point
$A$ to $B$, the disturbances travel from $A$ to $D$, i.e.
$R_{\rm{\gamma}}=AD$. Thus we have
$R_{\rm{\gamma}}/v_{\rm{d}}=R_{\rm{BLR}}/c$, and then
\begin{equation}
R_{\rm{\gamma}}=\frac{R_{\rm{BLR}}}{c}v_{\rm{d}}.
\end{equation}
In this case, the $\gamma$-rays will lead the lines. Combing
equations (7) and (8), we have
\begin{equation}
R_{\rm{\gamma}}=-\frac{c\langle\tau_{\rm{ob}}\rangle}{1+z}\frac{1}{\cos
\theta}.
\end{equation}
In this special case (hereafter Case D), $R_{\rm{\gamma}}\la
R_{\rm{BLR}}$ is expected from equation (8). From $\overline
\tau^{-}_{\rm{ob}}=-2.86$ years,
$\theta=12^{\rm{\circ}}$--$21^{\rm{\circ}}$ and equation (9), we
have $R_{\rm{\gamma}}\la R_{\rm{radio}}=0.77$--0.81 pc. These
estimated $R_{\rm{\gamma}}$ and $R_{\rm{BLR}}=0.83$ pc satisfy
$R_{\rm{\gamma}}\la R_{\rm{BLR}}$. This tests the correctness of
$R_{\rm{\gamma}}\la R_{\rm{BLR}}$ expected from equation (8). This
test confirms the reliability of the time lags estimated by the
ZDCF method. Those estimated $R_{\rm{radio}}$ in Case B contain
these $R_{\rm{radio}}$ estimated in Case D. Thus Case D is a
special Case B, and it is possible and reasonable. Combining
equations (7) and (8), one can also obtain
\begin{equation}
R_{\rm{BLR}}=-\frac{c\langle\tau_{\rm{ob}}\rangle}{1+z}\frac{c}{v_{\rm{d}}}\frac{1}{\cos
\theta}.
\end{equation}
From $\overline \tau^{-}_{\rm{ob}}=-2.86$ years,
$v_{\rm{d}}=0.9$--$0.995c$,
$\theta=12^{\rm{\circ}}$--$21^{\rm{\circ}}$ and equation (10), we
have $R_{\rm{BLR}}=0.77$--0.90 pc. These estimated values contain
the typical size of $R_{\rm{BLR}}=0.83$ pc. This confirms the
reliability of the time lags estimated by the ZDCF method.

\section{DISCUSSION AND CONCLUSIONS}
The positions of $\gamma$-ray--emitting regions are still an open
and controversial issue in the researches on blazars. Based on the
method proposed in section 2, we attempt to locate the emitting
positions of $\gamma$-rays within the second bumps in the
broad-band SEDs of blazars. In our previous works
\citep{b55,b56,b10}, the internal absorption for 10 GeV--1 TeV
$\gamma$-rays were used to constrain $R_{\rm{\gamma}}$,
independent of how the $\gamma$-rays are produced. Here, we try to
locate $R_{\rm{\gamma}}$, independent of the energies of
$\gamma$-rays from the SSC and EC processes. We find two emitting
regions, the inner one at sub-pc--pc scales from the central black
hole and the outer one around tens of pc scales. The outer one
satisfies $R_{\rm{\gamma}}\gg R_{\rm{BLR}}$ (Case C). The inner
one in Case D satisfies $R_{\rm{\gamma}}\la R_{\rm{BLR}}$. The
inner one in Case B mostly satisfies $R_{\rm{\gamma}}\la
R_{\rm{BLR}}$. At the same time, the inner one in Case B partly
satisfies $R_{\rm{\gamma}}> R_{\rm{BLR}}$, i.e.
$R_{\rm{BLR}}<R_{\rm{\gamma}}\la$ 2.62 pc.

It was suggested $R_{\rm{\gamma}}\la R_{\rm{BLR}}$ \citep{b33}.
\citet{b32} argued $R_{\rm{\gamma}}\la R_{\rm{BLR}}$ for powerful
blazars. \citet{b82} assumed $R_{\rm{\gamma}}< R_{\rm{BLR}}$ for
the VHE $\gamma$-rays in 3C 279. \citet{b56} and \citet{b10}
suggested that $R_{\rm{\gamma}}$ is within the BLR for 3C 279.
\citet{b37} modelled the SEDs of bright {\it Fermi\/} blazars, and
they found that the position of the jet dissipation region
$R_{\rm{diss}}$ is smaller than $R_{\rm{BLR}}$ for 53 out of 57
FSRQs. However, $R_{\rm{diss}}>R_{\rm{BLR}}$ for BL Lacs. They
used $R_{\rm{BLR}}=10^{17}L^{1/2}_{\rm{d,45}}$ cm to estimate
$R_{\rm{BLR}}$ for BL Lacs and FSRQs, where $L_{\rm{d,45}}$ is
accretion disc luminosity in units of $10^{45} \/\ \rm{erg \/\
s^{-1}}$. It is appropriate to use
$R_{\rm{BLR}}=10^{17}L^{1/2}_{\rm{d,45}}$ cm to estimate
$R_{\rm{BLR}}$ for FSRQs, but not for BL Lacs because this
relation is derived from the type 1 AGNs. Thus it should be
reliable that $R_{\rm{diss}}<R_{\rm{BLR}}$ for blazars. If the jet
dissipation region is equivalent to the $\gamma$-ray--emitting
region, $R_{\rm{diss}}<R_{\rm{BLR}}$ is equivalent to
$R_{\rm{\gamma}}<R_{\rm{BLR}}$. For 3C 273,
$R_{\rm{diss}}<R_{\rm{BLR}}$ \citep{b37}. These previous findings
are consistent with our results of $R_{\rm{\gamma}}\la
R_{\rm{BLR}}$ obtained in Cases B and D. It was also argued
$R_{\rm{\gamma}}> R_{\rm{BLR}}$ \citep{b54,b79}. This is
marginally consistent with $R_{\rm{BLR}}<R_{\rm{\gamma}}\la$ 2.62
pc obtained in Case B. These agreements confirm the reliability of
the method and assumptions.

\citet{b18} suggested for 3C 279 that $R_{\rm{\gamma}}\gg
R_{\rm{BLR}}$ for VHE $\gamma$-rays. It is recently advanced that
the bulk of the $\gamma$-rays is produced in regions of the jet at
distances of tens of pc from the central black hole
\citep[e.g.][]{b76,b59}. \citet{b8} predicted the existence of
large scale synchrotron X-ray jets in radio-loud AGNs, especially,
the X-ray jets are bright on 10 kpc scales in most red blazars and
red blazar-like radio galaxies. According to their predictions,
the large scale synchrotron X-ray jets can produce VHE
$\gamma$-rays by the SSC process. \citet{b97,b96} predicted the
hot spots in lobes and the knots in jets to be possible GeV--TeV
emitters. {\it Fermi\/}/LAT may resolve the large scale
$\gamma$-ray emitters than the nuclear emitters. These previous
findings support $R_{\rm{\gamma}}\gg R_{\rm{BLR}}$ as we obtain in
Case C. Also, $R_{\rm{\gamma}}\gg R_{\rm{BLR}}$ in Case C is not
inconsistent with $R_{\rm{\gamma}}> R_{\rm{BLR}}$ of \citet{b54}
and \citet{b79}. These confirm the reliability of our results.

Our previous works are applicable to the $\gamma$-rays emitted
from the regions in powerful blazars, where $R_{\rm{\gamma}}$ is
not much larger than $R_{\rm{BLR}}$ \citep{b55,b56,b10}. The
method proposed here can locate $R_{\rm{\gamma}}$ in the jet. The
inner emitting regions with $R_{\rm{\gamma}}\la R_{\rm{BLR}}$ are
likely the major contributor of the $\gamma$-rays below 10 GeV,
for that the $\gamma$-rays above 10 GeV are subject to the
photon-photon absorption due to the dense external soft photons at
the inner regions \citep[e.g.][]{b55,b56,b10}. The outer emitting
regions with $R_{\rm{\gamma}}\gg R_{\rm{BLR}}$ are likely the
major contributor of the $\gamma$-rays above 10 GeV, for that
these $\gamma$-rays are not subject to the photon-photon
absorption due to the thin external soft photons at the outer
regions. For these possible $\gamma$-ray emitters at large scales
of kpc--Mpc \citep{b8,b97,b96}, our works are not applicable.

The most prominent features on VLBI images of jets in radio-loud
AGNs are the radio core and bright knots in the jet \citep{b42}.
\citet{b50} investigated the relation between AGN $\gamma$-ray
emission and pc-scale radio jets. They identified the pc-scale
radio core as a likely location for both the $\gamma$-ray and
radio flares. A few hundreds of Schwarzschild radii, sub-pc-scale,
is the preferred jet position where most of the dissipation occurs
\citep{b36,b38,b37}. \citet{b77} suggested that the blazar
emission zone is located at pc-scale distances from the nucleus.
\citet{b33} suggested that $R_{\rm{\gamma}}$ is at sub-pc scales.
\citet{b11} also suggested a sub-pc $\gamma$-ray--emitting region
from the central black hole. These previous findings support
$R_{\rm{\gamma}}$ at sub-pc--pc scales. These $R_{\rm{\gamma}}$ of
sub-pc--pc scales are consistent with those $R_{\rm{\gamma}}$
obtained in Cases B and D. These sub-pc--pc scale $R_{\rm{radio}}$
obtained in Cases B and D are also consistent with the previous
findings of the blazar emission zone and the dissipation zone.
These agreements confirm the reliability of our results.

It is recently advanced that the bulk of the $\gamma$-rays is
generated in regions of the jet at distances of tens of pc from
the central black hole \citep[e.g.][]{b76,b59}. For Case C, we
obtain the outer emitting regions of
4.67--$30.81<R_{\rm{\gamma}}\la 9.43$--62.31 pc and
$R_{\rm{radio}}=9.43$--62.31 pc. These outer emitting regions are
comparable to the $\gamma$-ray--emitting regions at distances of
tens of pc. \citet{b81} found evidence of variability on
timescales of few hours from the 1.5 years {\it Fermi}/LAT light
curves of FSRQs 3C 454.3 and PKS 1510-089. They concluded that
significant variability on such short timescales disfavor the
scenario in which the bulk of the $\gamma$-rays is produced at
distances of tens of pc \citep[e.g.][]{b76,b59}. The previous
researches show that there are two possible $\gamma$-ray--emitting
regions, one inside or around the BLR and the other outside the
BLR. This paper gives the same results. However, the method cannot
discriminate between positive lags and negative lags on
observational grounds alone (at least not with the current data),
and the application discussed in the paper does not distinguish
between the two proposed scenarios. We expect this situation to
change with future data, perhaps longer line light curves, such as
10--15 years. The longer line light curves could give stronger
constraints on the coupling of the radio light curves with the
line ones.

For a given line, the relevant time lags generally decrease as
radio frequency increases from 5 to 37 GHz. The trend is likely
from the radiative cooling of relativistic electrons. \citet{b9}
deduced the synchrotron time lag formula (see equation 9 therein).
This formula can be expressed as in the observer's frame
\begin{equation}
\tau^{\rm{ob}}_{\rm{lag}}(\rm{yrs})=1492.6\frac{\sqrt{1+z}B^{-3/2}}{\sqrt{\delta}(1+D)}
\left(\nu^{-1/2}_{\rm{H}}-\nu^{-1/2}_{\rm{L}}\right),
\end{equation}
where $D$ is the "Compton dominance" \citep[see e.g.][]{b34}, $B$
is the magnetic field strength in units of gauss, $\delta$ is the
Doppler factor, and $\nu_{\rm{H}}$ and $\nu_{\rm{L}}$ in units of
$\rm{GHz}$ are high and low frequencies in the observer's frame,
respectively. For 3C 273, \citet{b34} obtained $B=8.9 \/\ \rm{G}$
and $\delta=6.5$. Because the radio light curves used to calculate
the ZDCFs span more than 20 years and the line light curves span
about 7.5 years, it is better to derive $D$ by using the ratio of
synchrotron to $\gamma$-ray average luminosity. $D$ is of the
order of magnitudes of 1 \citep{b85}. We can obtain
$\tau^{\rm{ob}}_{\rm{lag}}(\rm{yrs})=12(\nu^{-1/2}_{\rm{H}}-\nu^{-1/2}_{\rm{L}})$
if $D=1$ is adopted. The total cooling of both synchrotron and
$\gamma$-ray emission can lead to
$\tau^{\rm{ob}}_{\rm{lag}}(\rm{yrs})=6(\nu^{-1/2}_{\rm{H}}-\nu^{-1/2}_{\rm{L}})$.
We calculate the ZDCFs and time lags between the light curves of
5, 8, 15, 22 and 37 GHz. The high frequency variations lead the
low frequency ones. The measured time lags
$\tau^{\rm{ob}}_{\rm{lag}}$ and the relevant frequency differences
$\nu^{-1/2}_{\rm{H}}-\nu^{-1/2}_{\rm{L}}$ are presented in Fig. 6.
The observational data are well consistent with the prediction of
$\tau^{\rm{ob}}_{\rm{lag}}=6(\nu^{-1/2}_{\rm{H}}-\nu^{-1/2}_{\rm{L}})$
(see Fig. 6). This agreement confirms the origin of radiative
cooling for the time lags between the radio light curves used
here. \citet{b67,b68} also found frequency-dependent time delays
for strong outbursts in several other blazars. In Fig. 7, we
compare the lags $\tau^{\rm{ob}}_{\rm{lag}}$ with the differences
of $\Delta \tau_{\rm{cent}}$ between $\tau_{\rm{cent}}$ listed in
Table 1. The line of $\Delta
\tau_{\rm{cent}}=\tau^{\rm{ob}}_{\rm{lag}}$ is consistent with the
measured data points (see Fig. 7). This agreement confirms that
the trend, i.e. the lags for a given line generally decrease as
radio frequency increases, most likely results from the radiative
cooling of relativistic electrons.

In addition, there is another possibility that lower frequencies
probe larger radii in the jet, as synchrotron self-absorption is
important for increasingly high radius with decreasing radio
frequency. The synchrotron self-absorption coefficient
$\alpha_{\rm{\nu}}$ is $\alpha_{\rm{\nu}}\propto
\nu^{-(n+4)/2}N_{\rm{e}}$, where $N_{\rm{e}}$ is the electron
density and $n$ is the electron distribution index. For a
homogeneous blob with a radius of $r$, the synchrotron
self-absorption optical depth $\tau_{\rm{\nu}}$ is
$\tau_{\rm{\nu}}=r\alpha_{\rm{\nu}}\propto
r\nu^{-(n+4)/2}N_{\rm{e}}\propto r \nu^{-(n+4)/2}/r^3=
\nu^{-(n+4)/2}/r^2$. Thus the radio frequency $\nu$ can probe the
radius $r$ that scales as $r\propto \nu^{-(n+4)/4}$. Hence, lower
frequencies probe larger radii in the jet due to the synchrotron
self-absorption. The higher frequencies will escape earlier from
the blob and later the lower ones as the blob expands. Thus the
lower frequencies lag the higher ones, and the relevant time lags
$\tau_{\rm{lag}}$ are related to frequencies. The difference in
lags listed in Table 1 could originate from the synchrotron
self-absorption, and it scales with frequencies as
$\tau_{\rm{lag}}\propto r_{\rm{L}}-r_{\rm{H}}\propto
\nu_{\rm{L}}^{-(1+n/4)}-\nu_{\rm{H}}^{-(1+n/4)}$. The dependence
of $\tau_{\rm{lag}}$ on frequencies is different from that of
equation (11).
\begin{figure}
\begin{center}
\includegraphics[width=2.0 in,angle=-90]{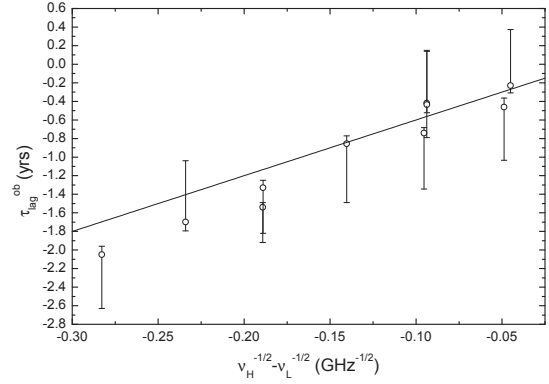}
 \end{center}
 \caption{Relation of time lags $\tau^{\rm{ob}}_{\rm{lag}}=t_{\rm{\nu_{H}}}-t_{\rm{\nu_{L}}}$ and frequency
 differences $\nu^{-1/2}_{\rm{H}}-\nu^{-1/2}_{\rm{L}}$
 between 37, 22, 15, 8 and 5 GHz. Solid line is the expectation from the radiative cooling.}
  \label{fig6}
\end{figure}
\begin{figure}
\begin{center}
\includegraphics[width=2.0 in,angle=-90]{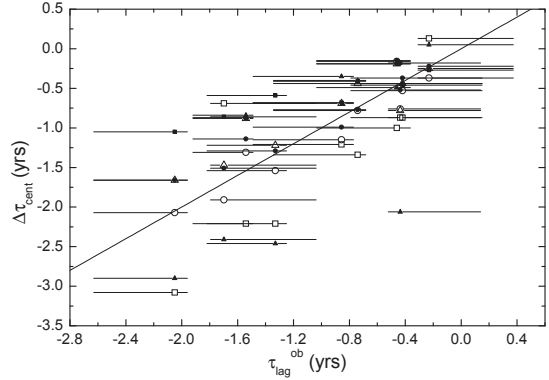}
 \end{center}
 \caption{$\Delta \tau_{\rm{cent}}$ vs $\tau^{\rm{ob}}_{\rm{lag}}$.
 Circles present H$\alpha$, squares H$\beta$ and triangles H$\gamma$.
 Open symbols present the negative lags and fulled symbols the positive lags in Table 1. Solid line is
 $\Delta \tau_{\rm{cent}}=\tau^{\rm{ob}}_{\rm{lag}}$.}
  \label{fig7}
\end{figure}

It seems possible to infer $R_{\rm{radio}}$ based on the time lags
of the radio synchrotron emission relative to the UV continuum
used by \citet{b62}. This approach seems more direct than based on
the lags of the broad lines relative to the radio emission. We
calculate the ZDCF between the light curves of the UV continuum
and the 37 GHz emission. There is only a little bump closer to the
zero-lag for the ZDCF and the little bump has
$r_{\rm{max}}=0.32\pm0.06$. In the ZDCFs between the Balmer lines
and this radio emission, the bumps used to calculate
$\tau_{\rm{cent}}$ have $r_{\rm{max}}=$ 0.6--0.7 that are much
higher than $r_{\rm{max}}=0.32\pm0.06$. This indicates that the
correlation of the UV continuum with this radio emission is much
weaker than the Balmer lines with this radio emission. For 22, 15,
8 and 5 GHz, there are the same cases as in 37 GHz. The UV
continuum is regarded as the ionizing continuum that drives the
broad lines through the photoionization process. Thus it is
expected that the correlation of the UV continuum with the radio
synchrotron emission should be more significant than the broad
lines with the radio emission. However, this expectation is
contrary to the measurements in the paper. This disagreement
indicates that the UV continuum is likely not the real ionizing
continuum. \citet{b62} argued that the UV continuum is much closer
to the ionizing continuum than the optical continuum used by
\citet{b46}. That is, the UV continuum is still not the real
ionizing continuum. Thus it is more reliable to derive
$R_{\rm{radio}}$ from the lags of the broad lines relative to the
radio emission than from those of the radio emission relative to
the UV continuum.

The relativistic shortening of variation timescales seems to have
significant effect on the the estimates of time lags. The
correlation between the ejection epochs of jet components
(superluminal radio knots) and the dips in the X-ray emission was
interpreted as accretion of the X-ray--emitting gas in the inner
accretion disc into the central black holes and ejection of a
portion of the infalling material into the jet \citep[see
e.g.][]{b58}. An instability in the accretion flow causing a
section of the inner disc break off. Part of this section is drawn
into the event horizon of the central black hole but with
considerable material and energy ejected down the jet. The loss of
this section of the inner disc causes a decrease in the soft X-ray
flux, which is observed as a dip. This disturbance from the inner
accretion disc to the jet is observed as the ejection of a
superluminal radio knot from the radio core of the jet. The dips
in the X-ray flux represent the onset of the disturbances in the
central engine. The knots represent the synchrotron emission of
the transported disturbances at the sites of ejections. Hence, the
intervals between the epochs of the dips $\Delta t_{\rm{dip}}$
represent those between the disturbances in the central engine.
The intervals between the ejection epochs of knots $\Delta
t_{\rm{knot}}$ result from the disturbances in the central engine.
In the radio galaxy 3C 120, it was found that there are 13
ejection epochs of radio knots with the corresponding dips in the
X-ray flux \citep{b22}. The ejection epochs $t_{\rm{knot}}$ and
the corresponding epochs of dips $t_{\rm{dip}}$ have a good
correlation with $t_{\rm{knot}}=t_{\rm{dip}}+0.19$. For the 13
data pairs of $t_{\rm{knot}}$ and $t_{\rm{dip}}$, $\Delta
t_{\rm{knot}}$ and $\Delta t_{\rm{dip}}$ between data pairs are
equal to each other within the uncertainties, i.e. $\Delta
t_{\rm{knot}}=\Delta t_{\rm{dip}}$. The correlation of
$t_{\rm{knot}}=t_{\rm{dip}}+0.19$ also gives $\Delta
t_{\rm{knot}}=\Delta t_{\rm{dip}}$. For the ejections of knots in
3C 120, there are the relevant local peaks in synchrotron emission
flux from the jet \citep{b22,b80a}. The intervals between the
epochs of peaks $\Delta t_{\rm{peak}}$ are equal to $\Delta
t_{\rm{knot}}$ of the relevant ejections within the uncertainties,
i.e. $\Delta t_{\rm{peak}}=\Delta t_{\rm{knot}}$. Combing $\Delta
t_{\rm{knot}}=\Delta t_{\rm{dip}}$ and $\Delta
t_{\rm{peak}}=\Delta t_{\rm{knot}}$, we have $\Delta
t_{\rm{peak}}=\Delta t_{\rm{dip}}$. The intervals $\Delta
t_{\rm{peak}}$ are measured by the light curves of beamed
synchrotron emission from the jet. The intervals $\Delta
t_{\rm{dip}}$ are measured by the light curves of un-beamed
emission from a disc--corona system \citep{b22}. These intervals
$\Delta t_{\rm{peak}}$ and $\Delta t_{\rm{dip}}$ are generated by
the same disturbances from the disc to the jet. Thus it is likely
that the relativistic effects on the time lags between them are
negligible as both variations of the beamed synchrotron emission
from the jet and the un-beamed emission from the disc--corona
system are mainly generated by the same disturbances from the disc
to the jet. In fact, the DCF method was employed to search the
time lag between the X-ray and the 37 GHz variations in 3C 120,
and one anti-correlation was found with the X-ray leading the
radio variations by $120\pm 30$ days \citep{b22}. By the DCF
method, \citet{b23a} obtained for 3C 273 that the UV light curve
leads the radio emission by a few months. Based on this lag,
\citet{b23} showed that the radio emission is located some 4 light
years from the central source along the jet. Though the
relativistic effects are not considered in these DCFs used to
estimate the time lags of the radio emission relative to the
un-beamed emission of the UV and the X-rays, receivable results
are obtained in these works. By analogy, it is likely that the
relativistic effects would not have significant influence on these
ZDCFs between the radio emission and the broad-line light curves
used in the paper. Hence, the inferred time lags from these ZDCFs
should not be affected by the relativistic effects very
significantly.

For testing the correctness of the time lags obtained by these
ZDCFs, we compare the 37 GHz light curve with the broad-line light
curves moved horizontally and vertically (see Fig. 8). For the
positive lags adopted, the line light curves are moved right by
2.8 years for H$\alpha$, 3.5 years for H$\beta$ and 4.0 years for
H$\gamma$. For the negative lags adopted, the line light curves
are moved left by 3.3 years for H$\alpha$, 3.2 years for H$\beta$
and 2.0 years for H$\gamma$. These moved line light curves are
basically co-varied with the radio light curve (see Fig. 8). The
relation of $\Delta t_{\rm{peak}}=\Delta t_{\rm{dip}}$ in 3C 120
also indicates that they should have similar observed timescales
if both variations of the beamed synchrotron emission from the jet
and the un-beamed emission from the disc--corona system are mainly
generated by the same disturbances from the disc to the jet. These
moved times are basically consistent with those time lags listed
in Table 1. The averages of these moved times for the positive and
negative lags are consistent with $\overline
\tau^{+}_{\rm{ob}}=3.20$ years and $|\overline
\tau^{-}_{\rm{ob}}|=2.86$ years obtained in these ZDCFs,
respectively. The local peaks in the radio light curve basically
have the corresponding ones in the moved line light curves. These
indicate that the relativistic effects would not have significant
influence on the estimates of time lags. The disturbances in the
accretion flow passing through the sites of the central ionizing
continuum could be imprinted on the ionizing continuum, the broad
emission lines and the disc-jet system. In order to simulate the
influence of the disturbances on the synchrotron emission from the
jet, the transporting process of the disturbances from the disc to
the jet should first be simulated by general relativistic
magnetohydrodynamics \citep[GRMHD, see e.g. ][]{b49a,b59a}. The
light curves of broad emission lines could be simulated by the
photoionization code CLOUDY \citep{b28a}. The GRMHD simulations
might give the site of radio synchrotron emission, and then it
will be easy to estimate the time lags of the radio emission
relative to the broad lines. Also, time lags can be derived from
those ZDCFs between the simulated light curves of radio and line
emission. Comparing the two kinds of time lags from simulations
should determine whether or not the relativistic effects are
considered in the ZDCF analysis between one beamed emission and
one un-beamed emission due to the same disturbances. These
simulations are out of the scope of this paper.
\begin{figure}
\begin{center}
\includegraphics[width=2. in,angle=-90]{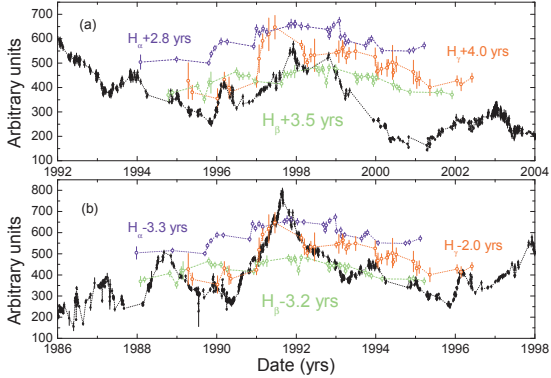}
 \end{center}
 \caption{The 37 GHz light curve versus the H$\alpha$, H$\beta$ and H$\gamma$ line light curves
 moved horizontally and vertically. $(a)$ the positive lags: moved right. $(b)$ the negative lags: moved left.}
  \label{fig8}
\end{figure}

In the paper, we propose a new method to derive the
$\gamma$-ray--emitting position $R_{\rm{\gamma}}$ from the time
lags $\tau_{\rm{ob}}$ of the $\gamma$-ray emission relative to the
broad lines (see Fig. 1). The method is also applicable to lower
energy bands, such as radio emission. $R_{\rm{\gamma}}$ depends on
four parameters $R_{\rm{BLR}}$, $v_{\rm{d}}$, $\tau_{\rm{ob}}$ and
$\theta$. As $\tau_{\rm{ob}}=0$, $\tau_{\rm{ob}}<0$ and
$\tau_{\rm{ob}}>0$ (Cases A, B and C), the broad lines zero-lag,
lag and lead the $\gamma$-rays, respectively. All cases are
unified into equation (7). The method is applied to FSRQ 3C 273.
Because the $\gamma$-ray light curves are very sparsely sampled
for 3C 273, it should be unreliable to employ them to estimate the
time lags. Fortunately, it was suggested that $R_{\rm{\gamma}}\la
R_{\rm{radio}}$ \citep{b24,b41,b50,b5}. Thus $R_{\rm{\gamma}}$
could be constrained by $R_{\rm{radio}}$ derived from the lags of
the radio emission relative to the broad lines. The ZDCF method is
used to analyze the correlations of the radio and infrared
emission with the broad lines H$\alpha$, H$\beta$ and H$\gamma$.
The broad lines lag or lead the 5, 8, 15, 22 and 37 GHz emission
(see Fig. 3). However, there is a lack of correlation between the
infrared emission and the broad lines (see Fig. 4). The measured
lags are on the order of years (see Table 1). The measured lags
for a given line generally decrease as radio frequency increases
(see Table 1). This trend most likely results from the radiative
cooling of relativistic electrons (see Fig. 7). The measured
negative lags have an average of $\overline
\tau^{-}_{\rm{cent}}=-2.86$ years for the 37 GHz emission relative
to the broad lines. From $\overline \tau^{-}_{\rm{ob}}=-2.86$
years, $R_{\rm{BLR}}=2.70$ $\rm{ly}$, $v_{\rm{d}}=0.9$--$0.995c$,
$\theta=12^{\rm{\circ}}$--$21^{\rm{\circ}}$ and equation (7), we
obtain $R_{\rm{radio}}=0.40$--2.62 pc (Case B). These estimated
$R_{\rm{radio}}$ contain the typical size of $R_{\rm{BLR}}=0.83$
pc, i.e. the radio emitting regions in Case B are around the BLR.
Thus $R_{\rm{\gamma}}\la 0.40$--2.62 pc for Case B and these inner
emitting regions mostly satisfy $R_{\rm{\gamma}} \la
R_{\rm{BLR}}$. For Case D, the special Case B, $R_{\rm{\gamma}}\la
R_{\rm{radio}}=0.77$--0.81 pc and $R_{\rm{\gamma}} \la
R_{\rm{BLR}}$. The measured positive lags have an average of
$\overline \tau^{+}_{\rm{cent}}=3.20$ years for the 37 GHz
emission relative to the broad lines. From $\overline
\tau^{+}_{\rm{ob}}=3.20$ years, $R_{\rm{BLR}}=2.70$ $\rm{ly}$,
$v_{\rm{d}}=0.9$--$0.995c$,
$\theta=12^{\rm{\circ}}$--$21^{\rm{\circ}}$ and equation (7), we
obtain $R_{\rm{radio}}=9.43$--62.31 pc $\gg R_{\rm{BLR}}$ (Case
C). Considering the zero-lag point and the constraint of
$R_{\rm{\gamma}} \la R_{\rm{radio}}$, we have
4.67--$30.81<R_{\rm{\gamma}}\la 9.43$--62.31 pc for Case C. These
outer emitting regions satisfy $R_{\rm{\gamma}}\gg R_{\rm{BLR}}$.
The dependence of $R_{\rm{\gamma}}$ and $R_{\rm{radio}}$ on
$v_{\rm{d}}$ and $\theta$ and the allowed intervals of
$R_{\rm{\gamma}}$ are presented in Fig. 5. $R_{\rm{\gamma}}$ and
$R_{\rm{radio}}$ increase as $v_{\rm{d}}$ increases, but decrease
as $\theta$ increases (see Fig. 5). The uncertainties of
$v_{\rm{d}}$ and $\theta$ result in the larger intervals of
$R_{\rm{\gamma}}$ and $R_{\rm{radio}}$. The positions of emitting
regions in Cases B and D are consistent with those at sub-pc--pc
scales suggested by \citet{b11}, \citet{b33}, \citet{b50} and
\citet{b77}, and also with the dissipative positions at
sub-pc-scale proposed by others \citep{b36,b38,b77,b37}. The
positions of emitting regions in Case C are comparable to those at
distances of tens of pc advanced in the recent researches
\citep[e.g.][]{b76,b59}. The relative positions of the BLR and
these emitting regions are also consistent with those previous
findings. From $\overline \tau^{-}_{\rm{ob}}=-2.86$ years,
$v_{\rm{d}}=0.9$--$0.995c$,
$\theta=12^{\rm{\circ}}$--$21^{\rm{\circ}}$ and equation (10), we
obtain $R_{\rm{BLR}}=0.77$--0.90 pc, which contain the typical
size of $R_{\rm{BLR}}=0.83$ pc. These agreements confirm the
reliability of the method and assumptions. These inner emitting
regions are likely the major contributor of the $\gamma$-rays
below 10 GeV, and those outer ones are likely the major
contributor of the $\gamma$-rays above 10 GeV. The method is also
applicable to BL Lacs, in which broad lines are detectable, but
ionizing continuum is undetectable. The BLR sizes of BL Lacs may
be constrained by equation (10).

\section*{Acknowledgments}
We are grateful to the anonymous referee for constructive comments
and suggestions leading to significant improvement of this paper.
H.T.L. thanks the West PhD project of the Training Programme for
the Talents of West Light Foundation of the CAS, and National
Natural Science Foundation of China (NSFC; Grant 10903025) for
financial support. J.M.B. thanks support of NSFC (Grant 10973034).
J.M.W. is supported by NSFC (Grant 10733010). J.M.B. and J.M.W.
thanks support of the 973 Program (Grant 2009CB824800).

\label{lastpage}

\end{document}